\documentclass[pra,aps,showpacs,superscriptaddress,amsmath,amssymb,floatfix,twocolumn]{revtex4-1}
\usepackage{graphicx}
\usepackage{color}
\usepackage{young}
\usepackage{youngtab}
\usepackage[colorlinks,bookmarks=false,citecolor=blue,linkcolor=red,urlcolor=blue]{hyperref}
\usepackage{amsfonts}
\usepackage{amsmath}
\usepackage{times}
\usepackage{amssymb}
\usepackage{changes}
\usepackage[utf8]{inputenc}

\DeclareMathOperator{\sign}{sign}

\newcommand{\grad}{\boldsymbol{\nabla}}

\renewcommand{\Xi}{g}

\begin{document}
\title{Rashba cavity QED: a route towards the superradiant quantum phase transition}
\date{\today} 

\author{Pierre Nataf}
\affiliation{Laboratoire de Physique et Mod\'elisation des Milieux Condens\'es, Universit\'e Grenoble Alpes and CNRS, 25 rue des Martyrs, 38042 Grenoble, France}
\author{Thierry Champel}
\affiliation{Laboratoire de Physique et Mod\'elisation des Milieux Condens\'es, Universit\'e Grenoble Alpes and CNRS, 25 rue des Martyrs, 38042 Grenoble, France}
\author{Gianni Blatter}
\affiliation{Institute for Theoretical Physics, ETH Zurich, 8093 Zurich, Switzerland}
\author{Denis M. Basko}
\affiliation{Laboratoire de Physique et Mod\'elisation des Milieux Condens\'es, Universit\'e Grenoble Alpes and CNRS, 25 rue des Martyrs, 38042 Grenoble, France}

\begin{abstract}
We develop a theory of cavity quantum electrodynamics for a 2D electron gas in
the presence of Rashba spin-orbit coupling and perpendicular static magnetic
field, coupled to spatially nonuniform multimode quantum cavity photon fields.
We demonstrate that  the lowest polaritonic frequency of the full Hamiltonian
can vanish for realistic parameters, achieving the Dicke superradiant quantum
phase transition. This singular behaviour originates from soft spin-flip
transitions possessing a non-vanishing dipole moment at non-zero wave vectors
and can be viewed as a magnetostatic instability.
\end{abstract}


\maketitle

\section{Introduction}

The Dicke model, describing an ensemble of identical two-level systems (matter
excitations) coupled to a single bosonic (cavity photon) mode, is a
prototypical model of cavity quantum electrodynamics (QED)~\cite{Dicke_1954}.
In the so-called ultrastrong coupling regime \cite{Ciuti_2005,FornDiaz_2019},
when the coupling strength (Rabi frequency) becomes comparable to the energy
splitting of the two-level system and that of the photon, the Dicke
model model was shown to exhibit the so-called superradiant quantum phase
transition (SQPT) towards a ground state characterised by a finite static
average of the photon field~\cite{Hepp_1973,Emary_2003}. To the best of our
knowledge, this phase transition has never been observed at equilibrium,
although ultra-strong coupling regime has been reached in a two-dimensional
electron gas (2DEG) in a semiconductor nanostructure placed in a cavity and
subject to a perpendicular static magnetic field, so that the matter
excitations were represented by the cyclotron resonance~\cite{Scalari_2012}.
Moreover, a softening of the lowest polaritonic excitation has been
recently observed in this system~\cite{Keller2018}.

The Dicke model has an intrinsic flaw: it must be obtained by a reduction of a
full microscopic model of some matter system coupled to the electromagnetic
field, and typically, the assumptions used to justify this reduction, break
down when the model is pushed to the ultrastrong coupling. As a consequence,
the SQPT is usually prevented by the so-called ``no-go'' theorems
\cite{polonais_1975, nataf_nogo_2010, Todorov_2012, Hayn_2012, polini_2012,
Bamba_2014, rousseau_2017, andolina_nogo}. They express the simple fact
that a physical system cannot respond to a static uniform vector potential
which can be simply removed by a gauge tranformation. On top of the
driven-dissipative scenario~\cite{Carmichael_PRA_2007,Keeling_PRA_2019}, which
has been successfully realized with cold atoms~\cite{esslinger_2010},
different suggestions have been proposed to circumvent ``no-go'' theorems at
equilibrium. These include systems with magnetic-dipole interactions due to the presence of cavity magnetic fields~\cite{Knight_PRA} or its circuit QED analog with an inductive coupling~\cite{nataf_2010,nataf_2011}
that can be of much larger magnitude~\cite{devoret_2007}. Notably, in the past
two decades it has been shown in several works for different physical systems
that upon a proper microscopic treatment, the mysterious SQPT assumes a more
familiar shape of a ferroelectric \cite{Keeling_2007, Vukics_2015} or an
excitonic insulator \cite{mac_donald_ncomm,Georges_PRL_2019} instability.  In
these studies, the crucial role of the Coulomb interaction has been pointed
out. In addition, the instability occurred at length scales much shorter than
the cavity size, thereby questioning the very role of the cavity.

\begin{figure}[!h]
\includegraphics[width=0.9\linewidth]{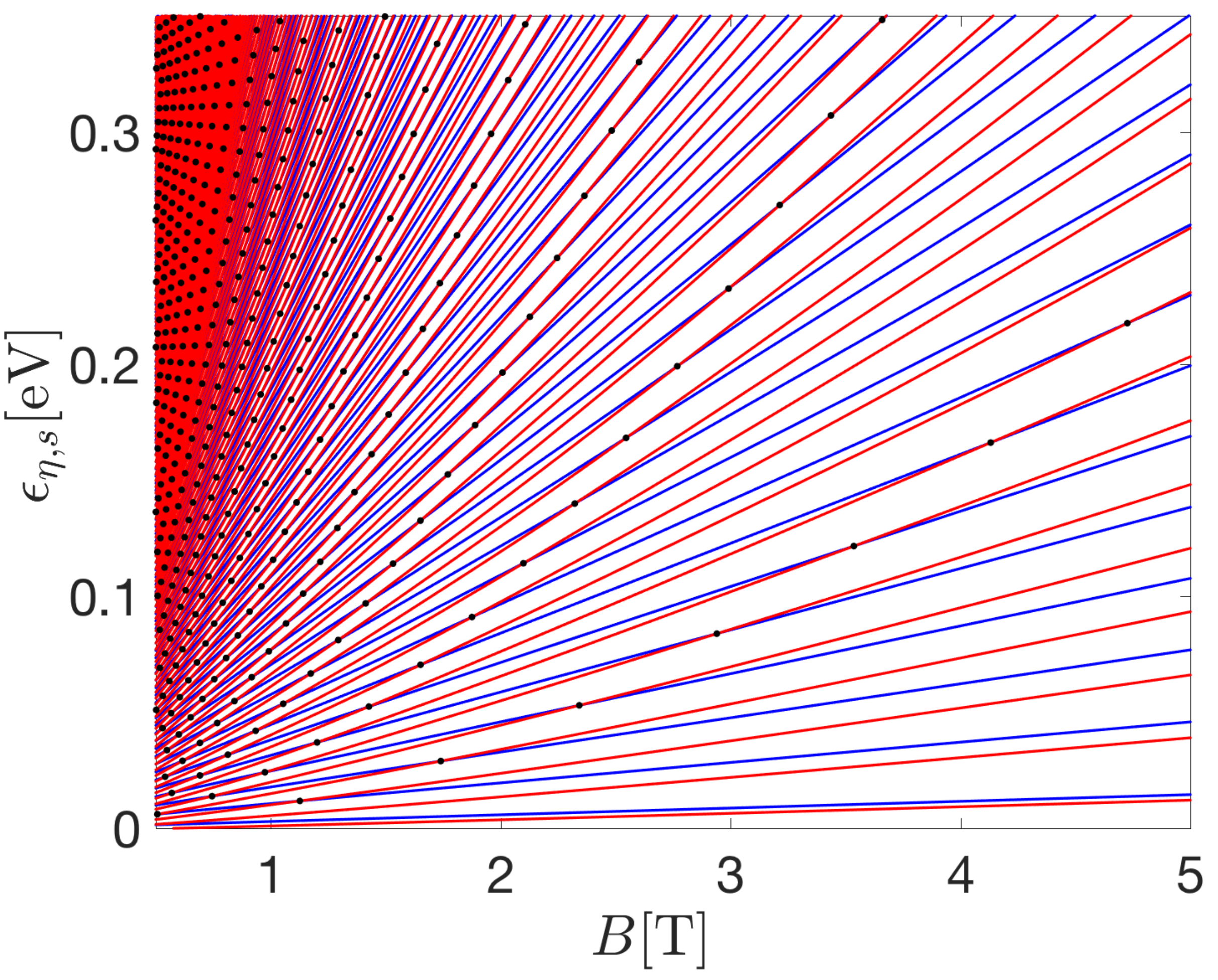}
\caption{\label{LL_figure_a40} 
(color online) LL energies $\epsilon_{\eta,s}$ from Eq.~\eqref{energy_Rashba}
versus $B$ in Tesla. The blue (red) curves correspond to $s=+1$ ($s=-1$). The
parameters are $\alpha=0.7\: \mbox{eV}\cdot\mbox{\AA}$ and $m^*=0.02m_0$. 
The black points mark the crossings where Eq. \eqref{equation_croisement} is
satisfied.}
%
\end{figure}

In this work, we present a model { without Coulomb interaction and still
exhibiting  a SQPT. Namely, we consider a 2DEG with Rashba spin-orbit
coupling, placed inside an optical cavity, and subject to a perpendicular
magnetic field~$B$. In the decoupled 2DEG, the Landau levels can cross at
certain values of~$B$ corresponding to dipole-allowed excitations with zero
energy. The presence of such intrinsic soft excitations greatly enhances the effect of the coupling to the transverse electromagnetic field.
We
develop a theory of Rashba cavity QED for integer filling factors and show
that this coupling  leads to
further softening of the system and appearance of some ``superradiant''
phases.  Crucially, the instability occurs at a finite wave vector of the
cavity field; to describe it, all high-energy cavity modes must be included
without any truncation in energy.  This instability is of magnetostatic
nature; the resulting ``superradiant'' phase is a remote relative of Condon
domains of spontaneous magnetization known since long ago for bulk metals in a
magnetic field~\cite{ShoenbergBook}.  Moreover, it turns out that this
instability can also occur without the cavity: the coupling to the free vacuum
field appears sufficient.

\section{The model}

It is well known \cite{Weisbuch_1992} that the effective strength of the light-matter coupling is enhanced if multiple copies of the material system are present.
We therefore consider $n_\mathrm{qw}$ identical quantum wells, each hosting a 2DEG with the single-electron Hamiltonian
containing a Rashba coupling term~\cite{rashba_1984},
\begin{align}
\mathcal{H}_{\text{2DEG}}={}&{}\frac{1}{2m^*}\left(\mathbf{p}+e\mathbf{A}/c\right)^2
+\alpha\left[\boldsymbol{\sigma} \times (\mathbf{p}+e\mathbf{A}/c)\right]_z.
\label{Rashba_Hamiltonian}
\end{align}
Here $\mathbf{p}=-i(\partial_x,\partial_y)$ is the 2D in-plane electron
momentum (we set $\hbar=1$), $m^*$~is the effective mass,
$\boldsymbol{\sigma}=(\sigma_x,\sigma_y,\sigma_z)$ is the vector of Pauli
matrices, and $\alpha$ is the Rashba spin-orbit coupling constant.  Typically, for some
existing InSb samples~\cite{Morgenstern_2010}, $m^*\simeq 0.02 m_0$ ($m_0$
being the free electron mass), $\alpha \simeq 0.7\: \mbox{eV}\cdot\mbox{\AA}$.
We assume $e>0$, the electron charge being $-e$.  Finally, the vector
potential $\mathbf{A}=\mathbf{A}^\mathrm{ext}+\mathbf{A}^\mathrm{cav}$
consists of two parts that we discuss separately.

First, $\mathbf{A}^\mathrm{ext}(\mathbf{r})=(-By,0,0)$ corresponds to the
external magnetic field~$B$, applied perpendicularly to the 2DEG plane (in the
$z$ direction). The resulting single-particle spectrum consists of Landau
levels (LLs) with energies
\begin{align}
\label{energy_Rashba}
\epsilon_{\eta,s}=\omega_c \left(\eta+\frac{s}{2}\sqrt{
1+8\eta\gamma^2}\right),
\end{align}
where $\eta=0,1,2,\ldots$ is the LL index, $s=\pm1$ for $\eta\geq 1$ and $s=1$
for $\eta=0$ is what remains of the spin index, $\omega_c=eB/(m^*c)$ is the
cyclotron frequency, and $\gamma\equiv\alpha m^*l_B$ with
$l_B\equiv(eB/c)^{-1/2}$ being the magnetic length. In Fig.\
\ref{LL_figure_a40}, we show the LL energies $\epsilon_{\eta,s}$ for
parameters consistent with InSb~\cite{Morgenstern_2010}; the spectrum exhibits
crossings that occur between  LLs $(\eta_1,s)$ and $(\eta_2,-s)$
satisfying the conditions \cite{Champel2013}} $\vert \eta_1-\eta_2\vert >1$ and
\begin{align}
\alpha^2=\frac{\eta_1+\eta_2-\sqrt{4\eta_1\eta_2+1}}{2(m^*l_B)^2}.\label{equation_croisement}
\end{align}
Note that levels with the same $s$ never cross.
Each Landau level has a degeneracy $L_xL_y/(2\pi{l}_B^2)$ where $L_xL_y$ is
the sample area. We assume to be at zero temperature, at  a fixed electron
density~$n_e$, and at  a magnetic field corresponding to an integer filling
factor $\nu\equiv2\pi{l}_B^2n_e$.  Indeed, the SQPT is associated with a
change in the character of the non-degenerate ground state of a gapped system.
Lifting the ground state degeneracy, which occurs at fractional fillings,
represents a totally different problem.
 

The  vector potential $\mathbf{A}^\mathrm{cav}(\mathbf{r})$ of the cavity
field is defined by the mode expansion, determined by the cavity shape. For
simplicity, here we consider a perfect metallic cavity whose dimensions
satisfy $L_x \gg L_z \gg L_y$, filled by a material with a dielectric
constant~$\varepsilon$. Then, one can consider only resonator modes with wave
vectors $\mathbf{q}=(q_x,0,q_z)$, where $q_x$ is varying continuously and
$q_z=\pi n_z/L_z$, $n_z=1,2,3,\ldots$ The corresponding  mode frequencies 
are $\omega^\mathrm{cav}_{q_x,n_z}=(c/\sqrt{\varepsilon})
\sqrt{q_x^2+q_z^2}$.  The cavity vector potential then reads
\cite{kakazu_kim,Hagenmuller_2010}
\begin{align}\label{eq:Acav}
\mathbf{A}^\mathrm{cav}(\mathbf{r})={}&{}\sum_{q_x,n_z}
\sqrt{\frac{4\pi}{L_xL_yL_z\varepsilon\omega_{\mathbf{q}n_z}}}\,
\mathbf{u}_y\sin\frac{n_z\pi{z}}{L_z}
\nonumber\\
{}&{}\times\left(a_{q_x,n_z}e^{iq_xx}+a_{q_x,n_z}^\dagger e^{-iq_x x}\right),
\end{align}
where $a_{q_x,n_z}^\dagger$ and $a_{q_x,n_z}$ are
the photon creation and annihilation operators and $\mathbf{u}_y$ is the unit
vector in the $y$ direction.  We assume the whole 2DEG sample with $n_{\mathrm{qw}}$ quantum wells to be much thinner than $L_z$  and
placed in the middle of the cavity. Then, what enters Eq.~(\ref{Rashba_Hamiltonian}), is
$\mathbf{A}_\mathrm{cav}(z=L_z/2)$ and the modes with even~$n_z$ are
decoupled.

\section{Polariton modes and instability}

The ``superradiant'' instability is signalled by the vanishing of the
lowest polariton frequency. To find the polariton modes -- the excitations of
the coupled 2DEG-cavity system -- one can proceed in several ways. For
example, similarly to that adopted in Ref.~\cite{Hagenmuller_2010} for the
same problem without spin-orbit coupling, one writes the 2DEG many-body
Hamiltonian in terms of creation and annihilation operators for inter-LL
excitations, which are approximately bosonic; then the full Hamiltonian of the
2DEG and the cavity becomes bilinear in the bosonic operators and thus can be
diagonalized by the appropriate Bogoliubov transformation. Alternatively, one
can write the (zero-temperature or Matsubara) action for coupled electron and
photon fields, integrate out the electrons, and expand the resulting bosonic
action to the second order in the photon vector potential. Both (rather
standard) calculations are given in the respective appendices A and B, and
their equivalence is checked explicitly.

As a result, the polariton frequencies are given by the solutions of the
equation
%
%
\begin{equation}
\frac{c}{2\pi}\sqrt{c^2q_x^2-\varepsilon\omega^2}
\coth\frac{L_z\sqrt{q_x^2-\varepsilon\omega^2/c^2}}{2}=-{Q_{yy}(q_x,\omega)}{},
\label{eq:dispersion}
\end{equation}
where $Q_{yy}(q_x,\omega)$ is the susceptibility determining the linear
response $j_y = Q_{yy}(\delta{A}_y/c)e^{iq_xx-i\omega{t}}$ of the 2D electron
current density $j_y$ to a perturbing vector potential $\delta\mathbf{A}$ on
top of $\mathbf{A}^\mathrm{ext}$ included in the unperturbed system,
$\mathbf{A}=\mathbf{A}^\mathrm{ext}+\delta\mathbf{A}e^{iq_xx-i\omega{t}}$.
The susceptibility consists of two contributions, the diamagnetic one and the
sum over all inter-LL transitions in all quantum wells,
%
%
\begin{equation}
Q_{yy}(q_x,\omega)=\frac{n_\mathrm{qw}n_e e^2}{m^*}\left[1
- \frac{\omega_c}{\nu}
\sum_{\ell \leq \nu<\ell'} \frac{\omega^\mathrm{LL}_{\ell'\ell}\,[\Xi^{\ell'\ell}_{q_x}]^2}{(\omega^\mathrm{LL}_{\ell'\ell})^2-\omega^2} \right].\label{eqQyy}
\end{equation}
Here the LL indices $(\eta,s)\equiv\ell$ are combined into a single label,
ordered according to the LL energies $\epsilon_\ell$,
Eq.~(\ref{energy_Rashba}), so that LLs with $\ell\leq\nu$ are filled, and
those with $\ell>\nu$ are empty. The transition energy
$\omega^\mathrm{LL}_{\ell'\ell}\equiv\epsilon_{\ell'}-\epsilon_\ell$, and the
reduced coupling constants $\Xi^{\ell'\ell}_{q_x}$ (dipole matrix
elements) are defined as
\begin{align} 
\label{coupling_rashba}
\Xi^{\ell'\ell}_{q_x} =&-\sqrt{2}\gamma \left( \sin{\theta_{\ell'}} \cos{\theta_{\ell}}\,\Theta^{\eta}_{\eta'-1}- \cos{\theta_{\ell'}} \sin{\theta_{\ell}}\,\Theta^{\eta-1}_{\eta'}\right) \nonumber\\
{}&{}+\sin{\theta_{\ell'}}\sin{\theta_{\ell}} \left( \sqrt{\eta -1}\,\Theta_{\eta'-1}^{\eta-2}- \sqrt{\eta }\,\Theta^{\eta}_{\eta'-1}  \right) \nonumber \\ 
{}&{}+ \cos{\theta_{\ell'}} \cos{\theta_{\ell}} 
\left( \sqrt{\eta}\,\Theta^{\eta-1}_{\eta'}- \sqrt{\eta+1}\,\Theta^{\eta+1}_{\eta'}  \right) ,
\end{align}
where $\tan{\theta}_{l}=[-1+s\sqrt{1+8 \eta \gamma^2}]/(\sqrt{8\eta} \gamma)$,
and the overlap function $\Theta_{n_2}^{n_1}$ containing the $q_x$ dependence
is given by 
\begin{align}
\Theta_{n_2}^{n_1}=
\label{theta}
\sqrt{\frac{m!}{M!}}\,e^{-\frac{\xi}{2}} \xi^{\frac{M-m}{2}}L^{M-m}_m(\xi)\,
S^{n_2-n_1},
 \end{align}
with $L^{M-m}_m(\xi)$ the generalized Laguerre polynomial of $\xi\equiv
l_B^2q_x^2/2$, $S=\sign\left[ q_x(n_2-n_1)\right]$, $m=\min\{n_1,n_2\}$, and
$M=\max\{n_1,n_2\}$.  Since at $q_x=0$ we have $\Theta_{n_2}^{n_1}=
\delta_{n_1,n_2}$, the reduced coupling constants $\Xi^{\ell'\ell}_{q_x=0}$
are non-zero only between consecutive LLs, $\eta'=\eta\pm1$, with no
restriction on~$s$. In contrast, at finite $q_x$, this selection rule is relaxed.

Equations~(\ref{eq:dispersion})--(\ref{theta}) represent the main analytical
result of this paper. Note that in Eq.~(\ref{eq:dispersion}) all information
about the cavity is on the left-hand side, while all information about the
2DEG is on the right.  At $\omega=0$ (i.e., at the sought
superradiant transition) the left-hand side is proportional to $c^2$ which is much larger than any velocity scale occurring in a typical solid.
Moreover, when $\omega=0$, the second term of the right-hand side of Eq.
\eqref{eqQyy} is nothing but the total sum of the oscillator strengths
$|g^{\ell'\ell}_{q_x}|^2/\omega^\mathrm{LL}_{\ell'\ell}$, which is the
fundamental quantum optics quantity that determines the occurrence of the SQPT
in multilevel systems coupled to cavity fields \cite{Hayn_2012}. It diverges at the level crossing, balancing the large $c^2$ factor in the left-hand side and allowing the existence of a solution to Eq.~(\ref{eq:dispersion}).

\begin{figure}[!h]
\centerline{\includegraphics[width=\linewidth]{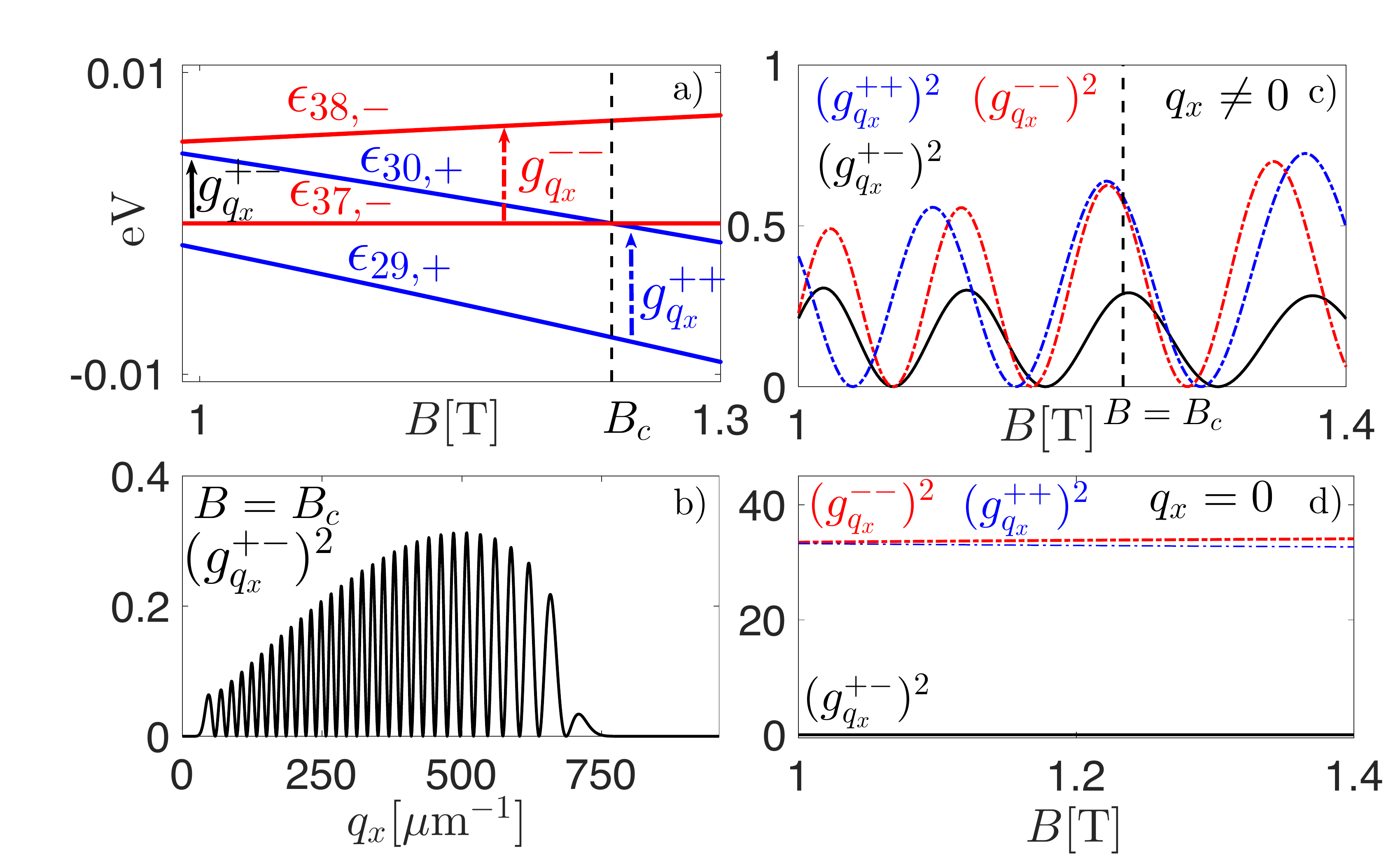}}
\caption{(color online)
a) LL energies $\epsilon_{29,+}$, $\epsilon_{30,+}$ $\epsilon_{37,-}$
$\epsilon_{38,-}$ in eV [see Eq. (\ref{energy_Rashba})] as a function of
$B$ (in Tesla) for the same parameters as in Fig \ref{LL_figure_a40}.  The gap
closes between the states $\vert 37,-\rangle$ and $\vert 30,+\rangle$ at $B_c
\simeq 1.23721$ T. In b), we
plot the square of the spin-flip dipole $(g^{+-}_{q_x})^2 \equiv
(g^{30+,37-}_{q_x})^2$ [ see Eq.~(\ref{coupling_rashba})] as a function
of $q_x$ at $B=B_c$. At $q_x=0$, it vanishes as a consequence of gauge
invariance. When $q_x$ increases, it oscillates and takes non-zero
values, opening the possibility of a diverging oscillator strength proportional to
$(g^{+-}_{q_x})^2/(\epsilon_{30,+}-\epsilon_{37,-})$. Finally, we plot $(g^{+-}_{q_x})^2,
(g^{++}_{q_x})^2$ and $(g^{--}_{q_x})^2$  as a function of $B$ for
$q_x=4\times 10^8\:\mbox{m}^{-1}$ in c) and $q_x=0$  in d).
\label{figure_couplage_4fois4} }
\end{figure}

The key reason is that for the spin-flip transitions at $q_x\neq0$,
the dipoles $\Xi^{\ell',\ell}_{q_x}$ can be non-zero even at a crossing
between $\epsilon_{\ell'}$ and $\epsilon_{\ell}$, opening the possibility of a
diverging oscillator strength. In Fig.\ \ref{figure_couplage_4fois4}, we present
an example of such crossing. This is in sharp contrast
with what happens at $q_x=0$, where gauge invariance demands the vanishing of
the dipole between eigenstates of equal energy (i.e., at the crossings of
energy levels), which is at the heart of many no-go theorems regarding the
Dicke SQPT for spatially uniform cavity fields \cite{polonais_1975,
nataf_nogo_2010,Todorov_2012,Hayn_2012,Bamba_2014,andolina_nogo}.

Indeed, for a generic single-electron Hamiltonian~$\mathcal{H}$, after the
minimal coupling replacement of the electron momentum $\mathbf{p}\rightarrow
\mathbf{p}+(e/c)\mathbf{A}_\mathrm{cav}$, where $\mathbf{A}_\mathrm{cav}$ is
the \emph{uniform} cavity field, the matrix element of the linear light-matter
coupling term between two arbitrary eigenstates $|1\rangle$, $|2\rangle$ of
$\mathcal{H}$, is proportional to that of the electron velocity
$\mathbf{v}=\partial\mathcal{H}/\partial\mathbf{p}=i[\mathcal{H},\mathbf{r}]$,
so that $\mathbf{v}_{12}=\langle 1 \vert [\mathcal{H},\mathbf{r}] \vert 2
\rangle =  (\epsilon_1-\epsilon_2) \langle 1 \vert \mathbf{r} \vert 2
\rangle$.  Then, at the crossing $\epsilon_1-\epsilon_2$ vanishes, and both
the velocity matrix element $\mathbf{v}_{12}$ and the oscillator strength
$\propto|\mathbf{v}_{12}|^2/(\epsilon_1-\epsilon_2)$ vanish. Crucially, this
argument does no apply to spatially non-uniform fields.

From a different perspective, gauge invariance imposes a constraint on
$Q_{yy}(q_x,\omega)$: a physical quantity (current) cannot respond to a static
spatially homogeneous vector potential, thus $Q_{yy}(0,0)=0$. This prohibits
the instability at $q_x=0$; models or approximations violating this constraint
can give wrong results.  $Q_{yy}(0,0)=0$ implies a cancellation between the
two terms in Eq.~(\ref{eqQyy}).  This cancellation is usually ensured by sum
rules such as the famous Thomas-Reiche-Kuhn 
strength of localized atomic systems \cite{polonais_1975, nataf_nogo_2010,
Todorov_2012, polini_2012, Hayn_2012, Bamba_2014, andolina_nogo}.  Here, we
have checked it numerically.

\begin{figure}
\centerline{\includegraphics[width=\linewidth]{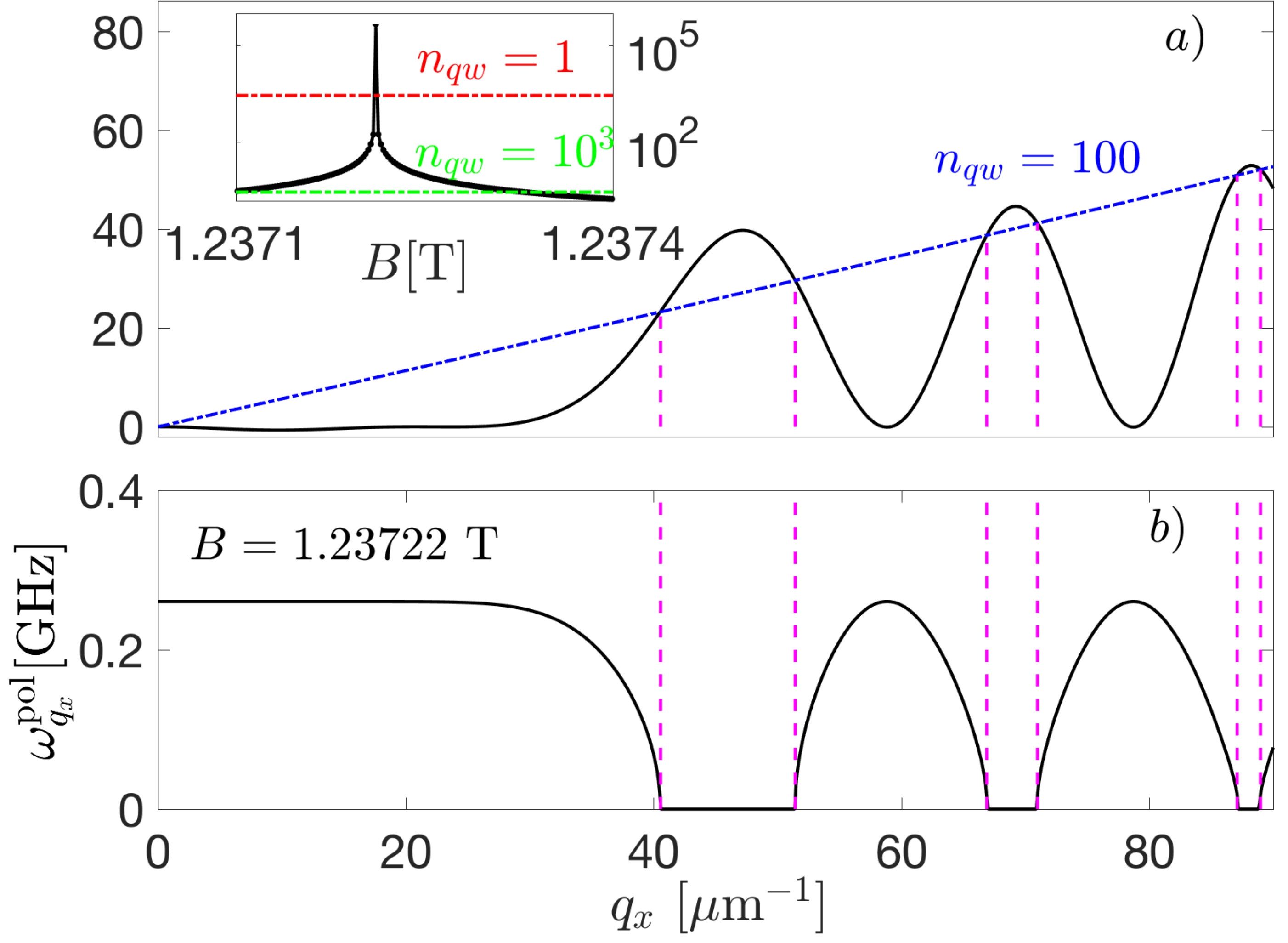}}
\caption{(color online) 
a)~The two sides of Eq. \eqref{eq:dispersion} at $\omega=0$ divided by
$(n_\mathrm{qw}n_ee^2/m^*)$ (the solid black curve for the right-hand side and
the dash-dotted blue one for the left-hand side, plotted versus $q_x$ (in $\mu\mbox{m}^{-1}$) for $B=1.23722\:\mbox{T}$
[which exceeds $B_c$, cf Fig.  \ref{figure_couplage_4fois4} (a)] for
$\alpha=0.7\: \mbox{eV}\cdot\mbox{\AA}$, $m^*=0.02\,m_0$ and $\nu=67$,
$L_z=20\:\mu\mbox{m}$, $\varepsilon=10$, $n_\mathrm{qw}=100$.  The crossings
of the two quantities give solutions of Eq. \eqref{eq:dispersion} for $q_x$ at
$\omega=0$ (vanishing polariton frequency).  In the inset, same quantities
plotted as a function of $B$ for $q_x=4.8 \times 10^7\:\mbox{m}^{-1}$, where
the red (resp.  green) correspond to  $n_\mathrm{qw}=1$ (resp.
$n_\mathrm{qw}=1000$).  At $B=1.23722\:\mbox{T}$, it happens at values of
$q_x$ shown by the vertical dashed lines for $n_\mathrm{qw}=100$. The lowest
polariton mode $\omega^\mathrm{pol}_{q_x}$, plotted in b), vanishes at these
values.}
\label{solutions}
\end{figure}

In Fig. \ref{solutions}(a), we plot the two sides of Eq.~(\ref{eq:dispersion})
at $\omega=0$ as functions of the cavity field wave-vector $q_x$ and show six
values of $q_x>0$ (shown in magenta dashed lines) when $\omega=0$ is a
solution. In Fig. \ref{solutions}(b), we display the lowest polariton
frequency $\omega_{q_x}^\mathrm{pol}$ which indeed vanishes at the indicated
values of~$q_x$.  Between two values of $q_x$ where Eq. \eqref{eq:dispersion}
is satisfied at $\omega=0$, the left-hand side of Eq.~(\ref{eq:dispersion}) is
smaller than the right-hand side and the system is in the ``superradiant''
phase.  In the inset of Fig. \ref{solutions}(a), we plot the two sides of
Eq.~(\ref{eq:dispersion}) at $\omega=0$ as a function of $B$ for a given value
of $q_x$, for $n_\mathrm{qw}=1$ and $n_\mathrm{qw}=10^3$. Essentially, we exploit a divergence
appearing around level crossings to make Eq.~(\ref{eq:dispersion}) have a solution, and then use large $n_\mathrm{qw}$ to have
an extended regime for the superradiant phase.

\begin{figure}
\centerline{\includegraphics[width=\linewidth]{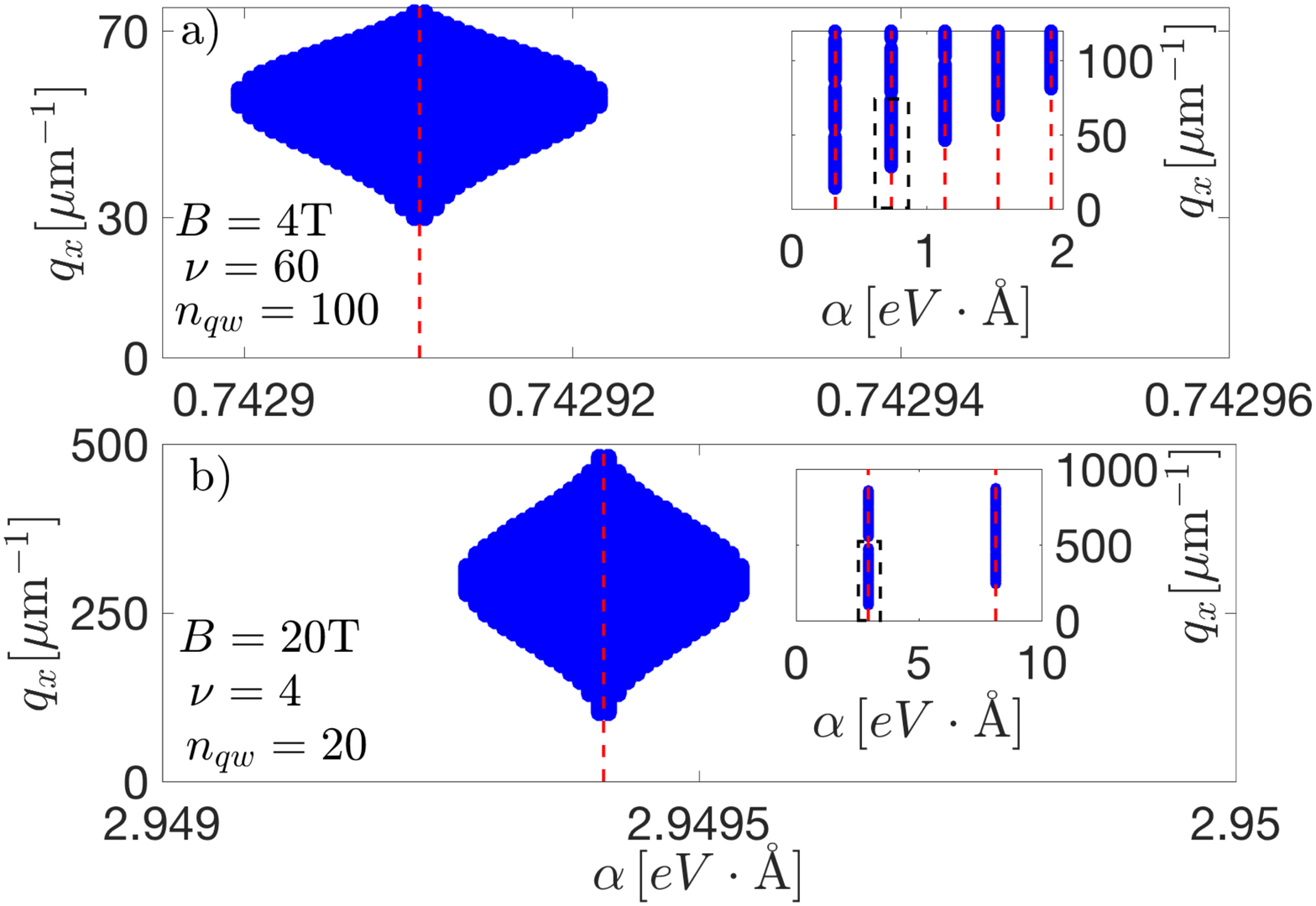}}
\caption{(color online)
Phase diagrams of the ``superradiant'' phase (in blue) in the parameter plane $(\alpha, q_x)$ 
 for different $B$, $\nu$ and $n_\mathrm{qw}$. 
 The inset display the phase diagrams for a larger range of parameters. Values of $\alpha$ for which there is an energy crossing
[cf Eqs. \eqref{energy_Rashba} and \eqref{equation_croisement}] are depicted in dashed red lines. The ``superradiant'' phases occur around these crossings.
a) $B=4\:\mbox{T}$, $\nu=60$ and $n_\mathrm{qw}=100$ ; b) $B=20\:\mbox{T}$, $\nu=4$ and $n_\mathrm{qw}=20$. 
}
\label{diagram}
\end{figure}

Fig. \ref{diagram} shows the phase diagram in the plane $(\alpha, q_x)$, for
fixed magnetic field $B$ and filling $\nu$.  The ``superradiant'' phase
appears for nonzero $q_x$ and for values of $\alpha$ very close to those given
by Eq.~(\ref{equation_croisement}) for some integers $\eta_1\neq \eta_2$
satisfying $\eta_1+\eta_2=\nu$, and depicted in dashed red lines in
Fig.~\ref{diagram}.  The characteristic width $\Delta\alpha$ of the
``superradiant'' regions on the phase diagram can be estimated as (see
Appendix~\ref{app:estimate})
\begin{equation}\label{eq:Deltaalpha}
\Delta\alpha\sim\frac{1}{\nu^2}\,\frac{n_\mathrm{qw}n_ee^2}{(m^*c)^2},
\end{equation}
and the typical scale of $q_x$ is given by the inverse cyclotron radius,
$(\sqrt\nu\,l_B)^{-1}$. The ``superradiant'' regions are very narrow; this
happens because the mechanism for the instability can be traced to the
magnetostatic interaction, as we discuss below. 
Typically, one arrives at the Dicke model assuming the light-matter coupling via the cavity electric field.
 However,  this electric field,  $\propto\partial\mathbf{A}_\mathrm{cav}
/\partial{t}$, vanishes at  $\omega=0$. The remaining magnetic interaction is intrinsically weak. These simple physical arguments are not obvious  from the equations.

From Eq. \eqref{eq:Deltaalpha} and Fig. \ref{diagram} we see  that small filling factors are favoring the ``superradiant'' phase.
This is in stark contrast to the condition of $\nu \gg 1$ formulated in Ref.~\cite{Hagenmuller_2010} to achieve the ultrastrong coupling regime. Again, the reason for this difference is that the SQPT obtained here is determined by the magnetic coupling and not the electric one.

\section{SQPT as a magnetostatic instability}

Consider the 2DEG in free space (no cavity). When placed in a static
homogeneous magnetic field $\mathbf{B}^\mathrm{ext}(\mathbf{r}) =
B\mathbf{u}_z$, it develops a static equilibrium magnetization
$\mathbf{M}^\mathrm{eq}(\mathbf{r})=M^\mathrm{eq}\mathbf{u}_z\,\delta(z-L_z/2)$.
Let us see at what conditions the system can spontaneously develop an
additional inhomogeneous magnetization $\delta\mathbf{M}(\mathbf{r})$ which
would produce an inhomogeneous magnetic field. Equivalently, we study the
stability of the described equilibrium configuration with respect to a small
perturbation, $\mathbf{B}=\mathbf{B}^\mathrm{ext} +
\delta\mathbf{B}(\mathbf{r})$.

We look for static magnetic field configurations
$\delta\mathbf{B}(\mathbf{r})$ which minimize the classical free energy
functional
\begin{align}\label{eq:Ftot=}
F[\delta\mathbf{B}]={}&{}
\int\left[\frac{(\mathbf{B}^\mathrm{ext}+\delta\mathbf{B})^2}{8\pi}
-\mathbf{M}^\mathrm{ext}\cdot\delta\mathbf{B}\right]d^3\mathbf{r}\nonumber\\
{}&{}+\int\left[-\mathbf{M}^\mathrm{eq}\cdot\delta\mathbf{B}
-\frac{1}{2}\,\chi_{kl}\,\delta{B}_k\delta{B}_l+\ldots\right]d^3\mathbf{r},
\end{align}
under the constraint $\grad\cdot\delta\mathbf{B}=0$. The first line contains
the energy density of the free field as well as the contribution due to
$\mathbf{M}^\mathrm{ext}$, the magnetization of the currents which
produce~$\mathbf{B}^\mathrm{ext}$. The second line is the free energy of the
2DEG.

Its quadratic part is determined by the static magnetic susceptibility
$\chi_{ij}(\mathbf{r},\mathbf{r}')$, which gives the response of the
magnetization $\delta\mathbf{M}$ to a magnetic field perturbation
$\delta\mathbf{B}$ on top of $\mathbf{B}^\mathrm{ext}$. Since
$\mathbf{B}=\grad\times\mathbf{A}$, and the static current density
$\mathbf{j}$ can be written in terms of the magnetization~$\mathbf{M}$ as
$\mathbf{j}=c\grad\times\mathbf{M}$, the susceptibility $\chi$ is related to
the current response $Q(\omega=0)$ found earlier on,
\begin{align}\label{eq:3D2Dreduction}
\chi_{zz}(\mathbf{r},\mathbf{r}')={}&{}
-\int\frac{dq_x}{2\pi}\,e^{iq_x(x-x')}
\frac{Q_{yy}(q_x,\omega=0)}{c^2q_x^2}\nonumber\\
{}&{}\times\delta(y-y')\,\delta(z-L_z/2)\,\delta(z'-L_z/2).
\end{align}
Here $\delta(y-y')$ appears because our starting model was restricted to 
a $y$-independent $\mathbf{A}_\mathrm{cav}(\mathbf{r})$,  see
Eq.~(\ref{eq:Acav}).

Varying $F+\int\lambda(\mathbf{r})\,\grad\cdot\mathbf{B}\,d^3\mathbf{r}$ with
the Lagrange multiplier $\lambda(\mathbf{r})$, we arrive at the equations
\begin{subequations}\begin{align}\label{eq:B=4piM}
&\mathbf{B}^\mathrm{ext}+\delta\mathbf{B}(\mathbf{r})=4\pi\mathbf{M}(\mathbf{r})
+\grad\int\frac{d^3\mathbf{r}'}{|\mathbf{r}-\mathbf{r}'|}\,\grad'\cdot\mathbf{M}(\mathbf{r}'),\\
\label{eq:M=chiB}
&M_k(\mathbf{r})={M}_k^\mathrm{ext}(\mathbf{r})
+{M}_k^\mathrm{eq}(\mathbf{r})
+\int\chi_{kl}(\mathbf{r},\mathbf{r}')\,\delta{B}_l(\mathbf{r}')\,d^3\mathbf{r}'.
\end{align}\end{subequations}
In Eq.~(\ref{eq:B=4piM}), the terms containing $\mathbf{B}^\mathrm{ext}$ and
$\mathbf{M}^\mathrm{ext}$ cancel out by construction; the terms with
$\mathbf{M}^\mathrm{eq}$ drop out because a magnetization
$\propto\mathbf{u}_z\,\delta(z-L_z/2)$ does not produce any magnetic field
except near the sample edges; thus, Eq.~(\ref{eq:B=4piM}) becomes a linear
homogeneous equation for $\delta\mathbf{B}(\mathbf{r})$. The existence
of a non-trivial solution corresponds to a direction along which the quadratic
form in Eq.~(\ref{eq:Ftot=}) is flat.

Let us try the modulation
\begin{equation}
\delta\mathbf{M}(\mathbf{r})=M_{q_x}\mathbf{u}_z\delta(z-L_z/2)\,e^{iq_xx},
\end{equation}
with some~$q_x>0$. The solution of Eq.~(\ref{eq:B=4piM}) for $\delta\mathbf{B}$,
\begin{align}
&\delta{B}_x(\mathbf{r})=-2i\pi{q}_xM_{q_x}e^{-q_x|z-L_z/2|-iq_xx}\sign(z-L_z/2),\nonumber\\
&\delta{B}_z(\mathbf{r})=2\pi{q}_xM_{q_x}e^{-q_x|z-L_z/2|-iq_xx},
\end{align}
when substituted in Eq.~(\ref{eq:M=chiB}), gives the magnetostatic instability
condition
\begin{equation}\label{eq:instability}
   -2\pi{q}_x\,\frac{Q_{yy}(q_x,\omega=0)}{c^2q_x^2}=1.
\end{equation}
This condition is identical to Eq.~(\ref{eq:dispersion}) at $\omega=0$ in the
limit $L_z\to\infty$, that is, in the absence of the cavity.

The obtained modulational instability of the magnetization is long known
for bulk paramagnetic materials in an external magnetic field, where it
gives rise to the so-called Condon domains of different
magnetization~\cite{ShoenbergBook}. Indeed, such a domain structure has the
characteristics of the ``superradiant'' phase: the magnetic field produced by
the domains corresponds to a non-zero expectation value of the photon field
$\langle{a}_{q_x,n_z}\rangle$. On the other hand, while the superradiant phase
in the Dicke model spontaneously breaks the $Z_2$ symmetry associated with the
sign of the spontaneously generated uniform cavity field~\cite{Emary_2003}
(variations of the model with higher symmetries have also been
considered~\cite{Hepp_1973,Nataf_2012,Baksic_2014}), the domain structure
obtained here spontaneously breaks at least the continuous translation
symmetry. In particular, for a sinusoidal modulation the $Z_2$ sign flip is
equivalent to translating the pattern by half a period. To make precise
statements about the broken symmetries, one would need a detailed quantitative
study of the  ``superradiant'' phase and of the resulting magnetization
profile; such a study is beyond the scope of the present paper.

\section{Conclusions and outlook}

We have proved that the SQPT can be reached in a cavity QED system with Rashba
spin-orbit coupling and non-uniform cavity resonator fields. Within our model, the appearance of the SQPT is a consequence of the singularity of the spin-flip transitions, for which the spin-flip dipole at the field wave-vector $q_x
\neq 0$ can be non-zero even if the transition energy vanishes.  Consequently,
the SQPT must occur close to LLs energy crossings (see
Figs.~\ref{solutions} and \ref{diagram}), and requires relatively fine tuning
of the magnetic field and electron density, as well as large
$n_\mathrm{qw}$.  We have shown that the SQPT can be viewed as a
magnetostatic instability of the 2DEG, the static cavity field corresponding to the magnetic field induced by a spatially modulated 2DEG magnetization.
  Moreover, the presence of the cavity
is not necessary : it can also happen via the coupling to the free vacuum field.

The microscopic model we studied in this paper has the minimal number of
ingredients necessary to produce SQPT. To make a connection with state of the
art experiments~\cite{Scalari_2012, Keller2018}, several other ingredients
must be introduced. Zeeman coupling of the 2DEG to the magnetic filed is
likely to enhance the effect since the spin contribution to the magnetic
susceptibility is usually paramagnetic. Coulomb interaction is also likely to
further soften the excitations due to the excitonic effect. A very
important ingredient is the disorder which lifts the LL degeneracy and
broadens the cyclotron resonance. Coherent state based methods have been quite
successful in describing the local density of states in a 2DEG with smooth
disorder in a strong magnetic field~\cite{Champel2009, Champel2013};  the
effect of smooth disorder on the inter-LL transitions remains an open
problem.  Effects of strain, as well as mixing between bands with different
spins (through, for instance, the multiband Luttinger $6\times6$
$\mathbf{k}\cdot\mathbf{p}$ model \cite{luttinger}), which both importantly
impact the amplitude and the nature of the Rashba coupling
\cite{Winkler_1995,Winkler_2000,Moriya_PRL}, could also be studied.  Finally,
adapting our calculations to some other cavities, like the Split Ring
Resonator (SSR) and Complementary SRR \cite{Maissen_2014}, where an
additionnal geometric factor can enhance the light matter interaction, could
also be useful. Finally, in this paper, we focused on the instability, leaving aside the study of the ``superradiant'' phase itself, a topic that
deserves future investigation as well. Moreover, the new physical ingredients
listed above, will also be important in determining the properties of the
``superradiant'' phase.


\acknowledgements

We acknowledge fruitful discussions with Tobias Wolf, Matteo Biondi, J\'erome
Faist, Giacomo Scalari, Janine Keller, Maksym Myronov, Rai Moriya, Takaaki
Koga,  Roland Winkler, Cristiano Ciuti and David Hagenm\"{u}ller.

\appendix

\section{Bogoliubov transformation approach}

The single-electron eigenstates of Hamiltonian~(\ref{Rashba_Hamiltonian}) with the external vector potential $\mathbf{A}_\mathrm{ext}=(-By,0,0)$ are labeled by $\ell\equiv(\eta,s)$ and $p_x$, the momentum in the $x$~direction, taken as an integer multiple of $2\pi/L_x$, with $L_x$ being the sample length the $x$~direction. It is convenient to order $\ell$ according to the energies $\epsilon_\ell$, given by Eq.~(\ref{energy_Rashba}). The spinor wave functions of the eigenstates are
\begin{align}
\psi_{\eta,s,p_x}(x,y)= \frac{e^{ip_xx}}{\sqrt{L_x}}
\begin{pmatrix} \sin{\theta_{\eta{s}}}\, \phi_{\eta-1}(y-p_xl_B^2) \\  i \cos\theta_{\eta{s}}\, \phi_{\eta}(y-p_xl_B^2) \end{pmatrix}, 
\label{wavefunction_rashba}
\end{align}
with $\tan{\theta}_{\eta{s}}=-u_\eta+s\sqrt{1+u_\eta^2}$, $u_\eta=1/(\sqrt{8\eta} \gamma)$ and the harmonic oscillator wave functions
\begin{align}
\phi_\eta(y)=\frac{e^{-y^2/(2l_B^2)}}{\sqrt{2^\eta \eta!\, l_B \sqrt{\pi}}} H_\eta(y/l_B),\end{align}
where $H_\eta$ is the Hermite polynomial of degree~$\eta$.
The many-body ground state of the 2DEG without coupling to the cavity is
\begin{align}
\vert F \rangle = \prod_{j=1}^{n_\mathrm{qw}}  \prod_{\ell \leq \nu}\prod_{p_x} c^{\dag}_{j,\ell,p_x} \vert 0 \rangle, \label{collective_GS_Rashba}
\end{align}
where $\vert 0 \rangle$ is the vacuum,  $c^{\dag}_{j,\ell,p_x}$ is the fermionic creation operator for an electron on the level~$\ell$ with momentum $p_x$ in the $j$th quantum well. The product over $p_x$ goes over all $L_xL_y/(2\pi{l}_B^2)$ allowed values.
Cavity photons induce collective electronic excitations which can be described by the approximatively bosonic bright modes creation operator, defined for $\ell,\ell'$ such that $\epsilon_\ell\leq \epsilon_{\nu}<\epsilon_{\ell'}$:
\begin{align}
b_{q_x,\ell'\ell}^{\dag}=
\sqrt{\frac{2\pi{l}_B^2}{n_\mathrm{qw}L_xL_y}}
\sum_{j,p_x} c^{\dag}_{j,\ell',p_x+q_x} c_{j,\ell,p_x}, 
\label{creation_rashba}
\end{align}
where the sum runs over the $n_\mathrm{qw}L_xL_y/(2\pi{l}_B^2)$ individual excitation between filled and empty LLs $\ell$ and $\ell'$, respectively. 

The full light-matter Hamiltonian $H$ contains four terms:
\begin{equation}
H=H_\mathrm{cav}+H_\mathrm{dia}+H_\mathrm{2DEG}+H_\mathrm{int}.\label{H}
\end{equation}
The free cavity photon part is
\begin{equation}
H_\mathrm{cav}=\sum_{q_x,n_z}  \omega^\mathrm{cav}_{q_x,n_z} a^{\dag}_{q_x,n_z} a_{q_x,n_z}.\end{equation}
The diamagnetic part of the light-matter interaction arises from the $\mathbf{A}_\mathrm{cav}^2$ term in the single-particle Hamiltonian~(\ref{Rashba_Hamiltonian}):
\begin{subequations}\begin{align}
&H_\mathrm{dia}=\sum_{q_x,n_z,n_z'}  D_{q_x}^{n_z,n_z'} a_{q_x,n_z}\left(a_{-q_x,n_z'}+a^{\dag}_{q_x,n_z'}\right)+\mbox{h. c.},\\
&D_{q_x}^{n_z,n_z'}= \frac{2\pi n_\mathrm{qw}n_e e^2}{\varepsilon{L}_zm^*\sqrt{\omega^\mathrm{cav}_{q_x,n_z}\omega^\mathrm{cav}_{q_x,n_z'}}}\,
\sin\frac{n_z\pi}2\sin\frac{n_z'\pi}2.
\end{align}\end{subequations}
$H_\mathrm{2DEG}$ stands for the electronic part of the Hamiltonian, which can be written using the operators in \eqref{creation_rashba}, as:
\begin{align}
H_\mathrm{2DEG}=\sum_{q_x}\sum_{\ell\leq\nu<\ell}
\omega^\mathrm{LL}_{\ell'\ell}\,b_{q_x,\ell'\ell}^{\dag} b_{q_x,\ell'\ell}.
 \end{align}
The corresponding energy differences $\omega^\mathrm{LL}_{\ell'\ell}=\epsilon_{\ell'}-\epsilon_\ell$ are associated to each transition across the Fermi level.
Finally, the linear in $\mathbf{A}_\mathrm{cav}$ term produces
\begin{equation}
H_\mathrm{int}=\sum_{q_x,n_z}\sum_{\ell\leq\nu<\ell'} i \Omega^{\ell'\ell}_{q_x,n_z} (a_{q_x,n_z}+a^{\dag}_{-q_x,n_z})\, b^{\dag}_{q_x,\ell'\ell}+\mbox{h.c.},
\end{equation}
where the $\ell,\ell'$ sum is again over $\epsilon_\ell\leq\mu\leq\epsilon_{\ell'}$, and the Rabi frequencies for each transition are determined by matrix elements of $e^{i q_x x} (-i\partial_y /m^*-\alpha \sigma_x)$ between different Landau level states:
\begin{align}
 \Omega^{\ell'\ell}_{q_x,n_z}&=\sqrt{n_\mathrm{qw}\frac{(e^2/c)}{\varepsilon \omega^\mathrm{cav}_{q_x,n_z}L_z/c}} \frac{\sin(n_z\pi/2)}{m^*l_B^2}\, \Xi^{\ell'\ell}_{q_x},
\end{align}
and the reduced coupling constants (dipole matrix elements) $\Xi^{\ell'\ell}_{q_x}$ are defined in Eq.~(\ref{coupling_rashba}).

In the following, we prove that the eigenfrequencies of the quadratic light-matter Hamiltonian $H$, shown in \eqref{H} can vanish for some wave vectors $q_x$.
Looking for the bosonic magnetopolariton modes in the form 
\begin{align}
&p_{q_x}=W_{q_x} a_{q_x} + Y_{q_x} a^{\dag}_{-q_x}\nonumber\\
&{}\qquad+\sum_{\ell\leq\nu<\ell'} 
\left(X_{q_x}^{\ell'\ell} b_{q_x,\ell'\ell} 
+ Z_{q_x}^{\ell'\ell}  b_{-q_x,\ell'\ell}^{\dag}\right),
\end{align}
which satisfies $[p_{q_x},H]=\omega_{q_x}^\mathrm{pol} p_{q_x}$, one has to diagonalize the corresponding Bogoliubov matrix 
$\mathcal{M}_{q_x}$.
The polaritonic frequencies $\omega_{q_x}^{pol}$ are the positive eigenvalues of $\mathcal{M}_{q_x}$, given by the solutions of the equation $\det(\mathcal{M}_{q_x}-\omega)=0$. The determinant can be written as
\begin{align}
&\det(\mathcal{M}_{q_x}-\omega)=
\left[1+ [1-\mathcal{I}_{q_x}(\omega)]\sum_{n_z\,\text{odd}}\frac{4D_{q_x}^{1,1} \omega^\mathrm{cav}_{q_x,1}}{(\omega^\mathrm{cav}_{q_x,n_z})^2-\omega^2}\right]\nonumber\\
&\qquad\qquad{}\times\prod_{\text{odd}\,n_z} \left[\omega^2-(\omega^\mathrm{cav}_{q_x,n_z})^2\right]
\prod_{\ell\leq \nu<\ell'}\left[\omega^2-(\omega^\mathrm{LL}_{\ell'\ell})^2\right],
\label{equation_det}
\end{align}
where we defined $\mathcal{I}_{q_x}(\omega)$ as
\begin{equation}
\mathcal{I}_{q_x}(\omega)= \frac{1}{\nu{m}^*l_B^2}
\sum_{\ell \leq \nu<\ell'} \frac{\omega^\mathrm{LL}_{\ell'\ell}\,[\Xi^{\ell'\ell}_{q_x}]^2}{(\omega^\mathrm{LL}_{\ell'\ell})^2-\omega^2}. \label{equation_mathcal_I}
\end{equation}
Using the fact that
\begin{equation}
\sum_{n_z\,\mathrm{odd}}\frac{1}{n_z^2+u^2}=\frac\pi{4u}\tanh\frac{\pi{u}}2,
\end{equation}
we can evaluate the $n_z$ sum in Eq.~(\ref{equation_det}) explicitly:
\begin{align}
&\sum_{n_z\,\text{odd}}\frac{4}{(\omega^\mathrm{cav}_{q_x,n_z})^2-\omega^2}={}\nonumber\\
&{}=\frac{\varepsilon{L}_z/c^2}{\sqrt{q_x^2-\varepsilon\omega^2/c^2}}
\tanh\frac{L_z\sqrt{q_x^2-\varepsilon\omega^2/c^2}}{2}.
\end{align}
Equating to zero the first prefactor in Eq.~(\ref{equation_det}), we obtain the equation for the polariton frequencies, the main analytical result of the present paper:
\begin{equation}
\coth\frac{L_z\sqrt{q_x^2-\varepsilon\omega^2/c^2}}{2}=\frac{e^2}c\,
\frac{2\pi n_\mathrm{qw}n_e}{m^*}\,
\frac{\mathcal{I}_{q_x}(\omega)-1}{\sqrt{c^2q_x^2-\varepsilon\omega^2}},
\label{eq:coth=}
\end{equation}
which is equivalent to Eqs.~(\ref{eq:dispersion}) and (\ref{eqQyy}) of the main text.

In the numerical calculations, we diagonalized the Bogoliubov matrix in a truncated basis of LLs and cavity modes, and checked for convergence. Typically, this required 50 LLs and $n_z$ up to $10^4$ for a given $q_x$.

\section{Effective action approach}

Let us describe the system by its Euclidean action $S$ (weight~$e^{-S}$) in the imaginary time $\tau$ varying in the interval $0\leq\tau\leq\beta$, where $\beta=1/T$ is the inverse temperature that we will eventually send to infinity ($T\to0$). The action consists of two parts, $S=S_\mathrm{cav}+S_\mathrm{2DEG}$. The action of the free cavity field is
\begin{equation}
S_\mathrm{cav}[\mathbf{A}^\mathrm{cav}]=\int\limits_0^\beta{d}\tau\int{d}^3\mathbf{r}
\left[\frac{\varepsilon(\partial_\tau\mathbf{A}^\mathrm{cav})}{8\pi{c}^2}
+\frac{[\grad\times\mathbf{A}^\mathrm{cav}]^2}{8\pi}\right].
\end{equation}
Here $\mathbf{A}^\mathrm{cav}$ is the vector potential of the cavity field in the Coulomb gauge, $\grad\cdot\mathbf{A}^\mathrm{cav}=0$, while $\varepsilon$ is the dielectric constant of the material filling the cavity. The electrons of the 2DEG, whose number is assumed to be fixed, are described by the action
\begin{equation}\label{eq:S2DEG=}
S_\mathrm{2DEG}[\psi^*,\psi,\mathbf{A}]=\int\limits_0^\beta{d}\tau\int{d}x\,dy\,
\psi^\dagger\left(\frac\partial{\partial\tau}+\mathcal{H}_\mathrm{2DEG}^{\otimes{n}_\mathrm{qw}}\right)\psi,
\end{equation}
where $\psi(x,y)$, $\psi^\dagger(x,y)$ are $2n_\mathrm{qw}$-component vectors of Grassmann variables, corresponding to two electron spin projections in each of the $n_\mathrm{qw}$ quantum wells. The single-electron Hamiltonian $\mathcal{H}_\mathrm{2DEG}$ is given by Eq.~(\ref{Rashba_Hamiltonian}) with $\mathbf{A}=\mathbf{A}^\mathrm{ext}+\mathbf{A}^\mathrm{cav}$. $\mathcal{H}_\mathrm{2DEG}$ a $2\times2$ matrix in the spin space, and in Eq.~(\ref{eq:S2DEG=}) it should be replicated $n_\mathrm{qw}$ times which is indicated by the superscript $\otimes{n}_\mathrm{qw}$.

Let us integrate out the electrons and obtain the effective action $S_\mathrm{eff}[\mathbf{A}^\mathrm{cav}]$ for the field only. It is convenient to pass to Matsubara frequencies
\begin{equation}
\mathbf{A}^\mathrm{cav}(\mathbf{r},\tau)=
T\sum_n\mathbf{A}^\mathrm{cav}(\mathbf{r},\omega_n)\,
e^{-i\omega_n\tau},
\end{equation}
where $\omega_n=2\pi{n}T$, and
$\mathbf{A}^\mathrm{cav}(-\omega_n)=[\mathbf{A}^\mathrm{cav}(\omega_n)]^*$.
Then the effective action can be written as a series in powers of~$\mathbf{A}^\mathrm{cav}$:
\begin{eqnarray}
&&S_\mathrm{eff}=
\frac{T}2\sum_{\omega_n}\int{d}^3\mathbf{r}
\left(\frac{\varepsilon\omega_n^2}{4\pi{c}^2}|\mathbf{A}^\mathrm{cav}|^2
+\frac{1}{4\pi}\,|\grad\times\mathbf{A}^\mathrm{cav}|^2\right)
\nonumber\\
&&\qquad{}-\frac{1}c\int{d}^3\mathbf{r}\,
\mathbf{j}^\mathrm{eq}(\mathbf{r})\cdot\mathbf{A}^\mathrm{cav}(\mathbf{r},\omega_n=0)
{}\nonumber\\
&&\qquad{}+\frac{T}{2c^2}
\sum_{\omega_n}\int{d}^3\mathbf{r}\,{d}^3\mathbf{r}'\,
\mathcal{Q}_{ij}(\mathbf{r},\mathbf{r}',i\omega_n){}\nonumber\\ &&\qquad\qquad{}\times
\,A_i^\mathrm{cav}(\mathbf{r},-\omega_n)\,
A_j^\mathrm{cav}(\mathbf{r}',\omega_n){}\nonumber\\
&&\qquad{}+\frac{T^2}{6c^3}\sum_{\omega_n,\omega_{n'}}
\int{d}^3\mathbf{r}\,{d}^3\mathbf{r}'\,{d}^3\mathbf{r}''\,
\mathcal{Q}^{(2)}_{ijk}(\mathbf{r},\mathbf{r}',\mathbf{r}'';i\omega_n, i\omega_{n'})
\nonumber\\ &&\qquad\qquad{}\times
A_i^\mathrm{cav}(\mathbf{r},-\omega_n-\omega_{n'})\,
A_j^\mathrm{cav}(\mathbf{r}',\omega_n)\,A_k(\mathbf{r}'',\omega_{n'}){}\nonumber\\
&&\qquad{}+\ldots,
\end{eqnarray}
where the kernels $\mathcal{Q}^{(n)}$ are given by the sum of electronic loops where the vertices $\psi^*\psi\mathbf{A}^\mathrm{cav}$ and $\psi^*\psi(\mathbf{A}^\mathrm{cav})^2$ appear $n_1$ and $n_2$ times, respectively, and $n_1+2n_2=n+1$. Generally speaking, the system is allowed to have an equilibrium current density, $\mathbf{j}^\mathrm{eq}(\mathbf{r})$, since it is placed in an external magnetic field and time reversal invariance is broken; in our specific case $\mathbf{j}^\mathrm{eq}=0$.

Analytical continuation in frequency of the kernels $\mathcal{Q}^{(n)}$ from the positive imaginary semiaxis $\omega=i\omega_n$, $\omega_n>0$, to the real axis $\omega$, gives the response functions which determine the response of the current density $\delta\mathbf{j}$ to a change in the vector potential, $\mathbf{A}=\mathbf{A}^\mathrm{ext}+\delta\mathbf{A}$; in the linear order,
\begin{equation}
\delta{j}_k(\mathbf{r},\omega)=-\int{d}^3\mathbf{r}'\,
\mathcal{Q}_{kl}(\mathbf{r},\mathbf{r}',\omega)\,\frac{\delta{A}_l(\mathbf{r}',\omega)}c.
\end{equation}
The polariton modes of the coupled system are determined by the quadratic part of~$S_\mathrm{eff}$ whose kernel is
\begin{align}
&\mathcal{D}^{-1}(\mathbf{r},\mathbf{r}',i\omega_n)
=\frac{\delta(\mathbf{r}-\mathbf{r}')}{4\pi}
\left[\varepsilon(\mathbf{r})\,\frac{\omega_n^2}{c^2}
+\grad\times\grad\times\right]\nonumber\\
&\hspace*{3cm}{}+\frac{1}{c^2}\,\mathcal{Q}_{ij}(\mathbf{r},\mathbf{r}',i\omega_n).
\end{align}
Namely, the polariton frequencies are solutions of the equation $\det\mathcal{D}^{-1}(\omega)=0$, where $\mathcal{D}^{-1}(\omega)$ is the analytical continuation of $\mathcal{D}^{-1}(i\omega_n)$ from the positive imaginary semiaxis and the determinant should be understood in the operator sense. Equivalently, the polariton eigenmodes are the nonzero solutions of
\begin{equation}\label{eq:Maxwell=}
\left[\varepsilon\,\frac{\omega^2}{c^2}+\nabla^2\right]A_i(\mathbf{r})=
\frac{4\pi}{c^2}\int\mathcal{Q}_{ij}(\mathbf{r},\mathbf{r}',\omega)\,A_j(\mathbf{r}')\,d^3\mathbf{r}',
\end{equation}
with $\grad\cdot\mathbf{A}=0$,
which is just the third Maxwell's equation with a source current.
The SQPT corresponds to the appearance of an unstable direction in the quadratic form with the kernel $\mathcal{D}^{-1}(\omega=0)$.

It should be noted that the zero-frequency limit of the Kubo susceptibility $\mathcal{Q}(\omega)$, which was obtained by the analytical continuation of the Matsubara susceptibility $\mathcal{Q}(i\omega_n)$ from the positive imaginary semiaxis $\omega_n>0$, is, generally speaking, different from the value of the Matsubara susceptibility $\mathcal{Q}(i\omega_n)$ taken directly at $\omega_n=0$. The Kubo limit $\mathcal{Q}(\omega\to{0})$ corresponds to evaluating the current with LL wave functions perturbed by a static $\delta\mathbf{A}$ and keeping the LL populations unperturbed; the Matsubara limit $\mathcal{Q}(i\omega_n=0)$ also takes into account the change in the LL populations which occurs because of the shift of the LL energies while the temperature and the chemical potential are kept fixed. In other words, they correspond to different order of limits $\omega\to0$ and population relaxation time to infinity. In our case, we are working at zero temperature and fixed electron density, rather than at fixed chemical potential. Thus, we take the Kubo limit.

By gauge invariance, a static vector potential $\mathbf{A}(\mathbf{r})$ can affect observable quantities only via the associated magnetic field $\mathbf{B}=\grad\times\mathbf{A}$. Also, at zero frequency the continuity equation for the current density requires $\grad\cdot\mathbf{j}=0$, so one can write $\mathbf{j}=c\grad\times\mathbf{M}$, where $\mathbf{M}$ is called magnetization.
Then it is convenient to express the response function $\mathcal{Q}_{ij}(\mathbf{r},\mathbf{r}',\omega=0)$ in terms of the static magnetic susceptibility $\chi_{ij}(\mathbf{r},\mathbf{r}')$, which determines the response of the magnetization $\delta\mathbf{M}$ to a magnetic field perturbation $\delta\mathbf{B}$ on top of $\mathbf{B}^\mathrm{ext}$. Then,
\begin{equation}
\frac{1}{c^2}\,\mathcal{Q}_{ij}(\mathbf{r},\mathbf{r}',\omega=0)=-e_{ikl}e_{jmn}
\frac{\partial^2\chi_{ln}(\mathbf{r},\mathbf{r}')}{\partial{x}_k\partial{x}_m'},
\end{equation}
where $e_{ikl}$ is the Levi-Civita antisymmetric tensor.

Now we proceed to evaluation of the response function $Q_{kl}(\mathbf{r},\mathbf{r}',\omega)$ for the 2DEG with the Rashba spin-orbit coupling under an external magnetic field. The single-electron 2D current density operator, corresponding to Hamiltonian~(\ref{Rashba_Hamiltonian}) for a single quantum well is given by
\begin{equation}
\mathbf{j}=-\frac{\partial\mathcal{H}_\mathrm{2DEG}}{\partial(\mathbf{A}/c)}
=-\frac{e}{m^*}\,\mathbf{p}-e\alpha[\mathbf{u}_z\times\boldsymbol{\sigma}]
-\frac{e^2}{m^*}\,\frac{\mathbf{A}}c,
\end{equation}
where $\mathbf{u}_{x,y,z}$, denotes the unit vectors in the respective directions.
As $\mathbf{A}=\mathbf{A}^\mathrm{ext}+\delta\mathbf{A}$, the term proportional to $\delta\mathbf{A}$ gives the diamagnetic contribution to the response, while the rest should be plugged in the Kubo formula.
We neglect the 2DEG thickness compared to all other length scales, so all quantum wells contribute additively and the response has the form
\begin{align}
\mathcal{Q}_{kl}(\mathbf{r},\mathbf{r}',\omega)={}&{}
\int\frac{d^2\mathbf{q}}{(2\pi)^2}\,e^{iq_x(x-x')+iq_y(y-y')}\,
Q_{kl}(\mathbf{q},\omega) \nonumber\\
{}&{}\times\delta(z-L_z/2)\,\delta(z'-L_z/2).
\end{align}
We are interested in the $Q_{yy}$ component at $\mathbf{q}=q_x\mathbf{u}_x$:
\begin{align}
Q_{yy}(q_x,\omega)={}&{}
\frac{2n_\mathrm{qw}}{L_xL_y}\sum_{\ell\leq\nu<\ell'}\sum_{p_x,p_x'}
\frac{\omega^\mathrm{LL}_{\ell'\ell}
|\langle\ell,p_x|j_y(q_x)|\ell',p_x'\rangle|^2}
{\omega^2-(\omega^\mathrm{LL}_{\ell'\ell})^2}\nonumber\\
{}&{}+\frac{n_\mathrm{qw}n_e e^2}{m^*},
\end{align}
where the first term comes from the Kubo formula and the last is the diamagnetic one. 

Evaluation of the current matrix elements between the LL eigenstates~(\ref{wavefunction_rashba}) gives
\begin{align}
&\langle\ell,p_x|{j}_y(q_x)|\ell',p_x'\rangle\equiv
e\int{dx}\,dy\, e^{-iq_xx}\,
\nonumber\\
&{}\times\left[\frac{i}{2m^*}\left(\psi_{\ell,p_x}^\dagger\!\frac{d\psi_{\ell'p_x'}}{dy}
-\frac{d\psi_{\ell,p_x}^\dagger}{dy}\psi_{\ell'p_x'}\right)
-\alpha\,\psi_{\ell,p_x}^\dagger\!\sigma_x\psi_{\ell'p_x'}\right]
\nonumber\\
&{}=-i\,\frac{e}{\sqrt{2}m^*l_B}\,\Xi^{\ell'\ell}_{q_x}\,\delta_{p_x',p_x+q_x},
\end{align}
where $\Xi^{\ell'\ell}_{q_x}$ is given by Eq.~(\ref{coupling_rashba}). As a result,
\begin{equation}
Q_{yy}(q_x,\omega)=\frac{n_\mathrm{qw}n_e e^2}{m^*}
\left[-\mathcal{I}_{q_x}(\omega)+1\right],
\end{equation}
where $\mathcal{I}_{q_x}(\omega)$ is given by Eq.~(\ref{equation_mathcal_I}).

Equation~(\ref{eq:Maxwell=}) becomes
\begin{align}
&\left(\varepsilon\,\frac{\omega^2}{c^2}
-q_x^2+\frac{d^2}{dz^2}\right)A_y(z)\nonumber\\
&{}=\frac{4\pi}{c^2}\,\delta(z-L_z/2)\,Q_{yy}(q_x,\omega)\,A_y(L_z/2),
\label{eq:Maxwellz=}
\end{align}
with the boundary conditions $A_y(0)=A_y(L_z)=0$. It has one family of solutions, $A_y(z)\propto\sin(\pi{n}_zz/L_z)$ with even $n_z$, corresponding to the cavity modes which remain decoupled from the 2DEG since $A_y(L_z/2)=0$. The other family can be represented as\begin{equation}
A_y(z)\propto\sinh\left(\kappa||z-L_z/2|-L_z/2|\right),\quad
\kappa\equiv\sqrt{q_x^2-\varepsilon\,\frac{\omega^2}{c^2}}.
\end{equation}
It has a jump of the derivative at $z=L_z/2$, which should be matched to the right-hand side of Eq.~(\ref{eq:Maxwellz=}), yielding
\begin{equation}
-2\kappa\cosh\frac{\kappa{L}_z}2=\frac{4\pi}{c^2}\,{Q}_{yy}(q_x,\omega)
\sinh\frac{\kappa{L}_z}2,
\end{equation}
which is equivalent to Eq.~(\ref{eq:dispersion}) of the main text and Eq.~(\ref{eq:coth=}).

\section{Estimate for the ``superradiant'' phase width}
\label{app:estimate}

Let us choose some LL crossing, $\epsilon_{\eta_1,+}=\epsilon_{\eta_2,-}$, and focus on $\alpha$ close to that given by Eq.~(\ref{equation_croisement}) which is denoted by $\alpha_\mathrm{c}$. The LL indices $\eta_1,\eta_2$ satisfy $\eta_1+\eta_2=\nu$, and we denote $\eta_1-\eta_2=\delta$. We will assume $\nu\gg{1}$, and the order-of-magnitude estimate is expected to be valid for $\nu\sim{1}$ as well. The dimensionless Rashba coupling strength corresponding to the crossing is then
\begin{equation}
\gamma_\mathrm{c}\equiv\alpha_\mathrm{c}m^*l_B\approx\sqrt{\frac{\delta^2-1}{4\nu}}\ll{1}.
\end{equation}
When $\alpha$ is detuned away from the crossing, keeping both the magnetic field and the filling constant, the LL energy difference becomes
\begin{equation}
\omega^\mathrm{LL}=|\epsilon_{\eta_1,+}-\epsilon_{\eta_2,-}|
\approx\sqrt{4\nu\,\frac{\delta^2-1}{\delta^2}}\,\frac{|\alpha-\alpha_\mathrm{c}|}{l_B}.
\end{equation}
To estimate the coupling strength $g_{q_x}$ at the crossing, we use the well-known asymptotic expression for the generalized Laguerre polynomials with large index in terms of the Bessel function of the first kind, $J$, which gives
\begin{equation}
\Theta^{n_1}_{n_2}\approx J_{|n_1-n_2|}\!\left(\sqrt{2(n_1+n_2)\xi}\right).
\end{equation}
Next, we need to find the angles $\theta_{\eta_1,+}$ and $\theta_{\eta_2,-}$. To the leading order in $1/\sqrt\nu$,
\begin{equation}
\tan\theta_{\eta_1,+}=-\cot\theta_{\eta_2,-}=\sqrt{\frac{|\delta|-1}{|\delta|+1}},
\end{equation}
and thus $\theta_{\eta_2,-}=\theta_{\eta_1,+}+\pi/2+O(1/\nu)$. As a result, the leading terms in the coupling $g_{q_x}$ cancel. This happens because of approximate orthogonality of the spin part of the wave functions for $|\eta_1-\eta_2|\ll\eta_1,\eta_2$ and $s_1=-s_2$.

Expansion to the next order will generate many terms; the resulting bulky expressions are not very informative. To obtain an order-of-magnitude estimate, it is sufficient to notice that the coupling strength can be written in the form
\begin{equation}
|g_{q_x}^{+-}|^2=\frac{1}\nu\,\mathcal{G}_\delta\!\left(\sqrt\nu{q}_xl_B\right),
\end{equation}
where for $\delta\sim{1}$ the function $\mathcal{G}_\delta(x)$ decays faster than $x$ at $x\to{0}$, reaches the first maximum at $x\sim{1}$, and the width of this maximum is also $\sim{1}$. Then, assuming that $Q_{yy}(q_x,\omega=0)$ is dominated by a single term corresponding to the LL crossing under consideration and setting $\coth(q_xL_z/2)\to{1}$, we can write Eq.~(\ref{eq:dispersion}) of the main text as
\begin{equation}
\Lambda\sqrt\nu{q}_xl_B=\mathcal{G}_\delta\!\left(\sqrt\nu{q}_xl_B\right),\quad
\Lambda\sim|\alpha-\alpha_\mathrm{c}|\,\frac{(m^*c)^2\nu^2}{n_\mathrm{qw}n_ee^2}.
\end{equation}
This equation has no solutions for $q_x$ if $\Lambda\gg{1}$; solutions appear when $\Lambda\sim{1}$ [Eq.~(\ref{eq:Deltaalpha}) of the main text]; then the typical scale of $q_x$ and the width of the unstable $q_x$~interval are determined by $\sqrt\nu{q}_xl_B\sim{1}$.

\bibliographystyle{apsrev4-1}
\bibliography{SPT_Dicke.bib}

\begin{thebibliography}{41}%
\makeatletter
\providecommand \@ifxundefined [1]{%
 \@ifx{#1\undefined}
}%
\providecommand \@ifnum [1]{%
 \ifnum #1\expandafter \@firstoftwo
 \else \expandafter \@secondoftwo
 \fi
}%
\providecommand \@ifx [1]{%
 \ifx #1\expandafter \@firstoftwo
 \else \expandafter \@secondoftwo
 \fi
}%
\providecommand \natexlab [1]{#1}%
\providecommand \enquote  [1]{``#1''}%
\providecommand \bibnamefont  [1]{#1}%
\providecommand \bibfnamefont [1]{#1}%
\providecommand \citenamefont [1]{#1}%
\providecommand \href@noop [0]{\@secondoftwo}%
\providecommand \href [0]{\begingroup \@sanitize@url \@href}%
\providecommand \@href[1]{\@@startlink{#1}\@@href}%
\providecommand \@@href[1]{\endgroup#1\@@endlink}%
\providecommand \@sanitize@url [0]{\catcode `\\12\catcode `\$12\catcode
  `\&12\catcode `\#12\catcode `\^12\catcode `\_12\catcode `\%12\relax}%
\providecommand \@@startlink[1]{}%
\providecommand \@@endlink[0]{}%
\providecommand \url  [0]{\begingroup\@sanitize@url \@url }%
\providecommand \@url [1]{\endgroup\@href {#1}{\urlprefix }}%
\providecommand \urlprefix  [0]{URL }%
\providecommand \Eprint [0]{\href }%
\providecommand \doibase [0]{http://dx.doi.org/}%
\providecommand \selectlanguage [0]{\@gobble}%
\providecommand \bibinfo  [0]{\@secondoftwo}%
\providecommand \bibfield  [0]{\@secondoftwo}%
\providecommand \translation [1]{[#1]}%
\providecommand \BibitemOpen [0]{}%
\providecommand \bibitemStop [0]{}%
\providecommand \bibitemNoStop [0]{.\EOS\space}%
\providecommand \EOS [0]{\spacefactor3000\relax}%
\providecommand \BibitemShut  [1]{\csname bibitem#1\endcsname}%
\let\auto@bib@innerbib\@empty
\bibitem [{\citenamefont {Dicke}(1954)}]{Dicke_1954}%
  \BibitemOpen
  \bibfield  {author} {\bibinfo {author} {\bibfnamefont {R.~H.}\ \bibnamefont
  {Dicke}},\ }\href {\doibase 10.1103/PhysRev.93.99} {\bibfield  {journal}
  {\bibinfo  {journal} {Phys. Rev.}\ }\textbf {\bibinfo {volume} {93}},\
  \bibinfo {pages} {99} (\bibinfo {year} {1954})}\BibitemShut {NoStop}%
\bibitem [{\citenamefont {Ciuti}\ \emph {et~al.}(2005)\citenamefont {Ciuti},
  \citenamefont {Bastard},\ and\ \citenamefont {Carusotto}}]{Ciuti_2005}%
  \BibitemOpen
  \bibfield  {author} {\bibinfo {author} {\bibfnamefont {C.}~\bibnamefont
  {Ciuti}}, \bibinfo {author} {\bibfnamefont {G.}~\bibnamefont {Bastard}}, \
  and\ \bibinfo {author} {\bibfnamefont {I.}~\bibnamefont {Carusotto}},\ }\href
  {\doibase 10.1103/PhysRevB.72.115303} {\bibfield  {journal} {\bibinfo
  {journal} {Phys. Rev. B}\ }\textbf {\bibinfo {volume} {72}},\ \bibinfo
  {pages} {115303} (\bibinfo {year} {2005})}\BibitemShut {NoStop}%
\bibitem [{\citenamefont {Forn-D\'{\i}az}\ \emph {et~al.}(2019)\citenamefont
  {Forn-D\'{\i}az}, \citenamefont {Lamata}, \citenamefont {Rico}, \citenamefont
  {Kono},\ and\ \citenamefont {Solano}}]{FornDiaz_2019}%
  \BibitemOpen
  \bibfield  {author} {\bibinfo {author} {\bibfnamefont {P.}~\bibnamefont
  {Forn-D\'{\i}az}}, \bibinfo {author} {\bibfnamefont {L.}~\bibnamefont
  {Lamata}}, \bibinfo {author} {\bibfnamefont {E.}~\bibnamefont {Rico}},
  \bibinfo {author} {\bibfnamefont {J.}~\bibnamefont {Kono}}, \ and\ \bibinfo
  {author} {\bibfnamefont {E.}~\bibnamefont {Solano}},\ }\href {\doibase
  10.1103/RevModPhys.91.025005} {\bibfield  {journal} {\bibinfo  {journal}
  {Rev. Mod. Phys.}\ }\textbf {\bibinfo {volume} {91}},\ \bibinfo {pages}
  {025005} (\bibinfo {year} {2019})}\BibitemShut {NoStop}%
\bibitem [{\citenamefont {Hepp}\ and\ \citenamefont {Lieb}(1973)}]{Hepp_1973}%
  \BibitemOpen
  \bibfield  {author} {\bibinfo {author} {\bibfnamefont {K.}~\bibnamefont
  {Hepp}}\ and\ \bibinfo {author} {\bibfnamefont {E.~H.}\ \bibnamefont
  {Lieb}},\ }\href {\doibase https://doi.org/10.1016/0003-4916(73)90039-0}
  {\bibfield  {journal} {\bibinfo  {journal} {Annals of Physics}\ }\textbf
  {\bibinfo {volume} {76}},\ \bibinfo {pages} {360 } (\bibinfo {year}
  {1973})}\BibitemShut {NoStop}%
\bibitem [{\citenamefont {Emary}\ and\ \citenamefont
  {Brandes}(2003)}]{Emary_2003}%
  \BibitemOpen
  \bibfield  {author} {\bibinfo {author} {\bibfnamefont {C.}~\bibnamefont
  {Emary}}\ and\ \bibinfo {author} {\bibfnamefont {T.}~\bibnamefont
  {Brandes}},\ }\href {\doibase 10.1103/PhysRevE.67.066203} {\bibfield
  {journal} {\bibinfo  {journal} {Phys. Rev. E}\ }\textbf {\bibinfo {volume}
  {67}},\ \bibinfo {pages} {066203} (\bibinfo {year} {2003})}\BibitemShut
  {NoStop}%
\bibitem [{\citenamefont {Scalari}\ \emph {et~al.}(2012)\citenamefont
  {Scalari}, \citenamefont {Maissen}, \citenamefont {Tur{\v c}inkov{\'a}},
  \citenamefont {Hagenm{\"u}ller}, \citenamefont {De~Liberato}, \citenamefont
  {Ciuti}, \citenamefont {Reichl}, \citenamefont {Schuh}, \citenamefont
  {Wegscheider}, \citenamefont {Beck},\ and\ \citenamefont
  {Faist}}]{Scalari_2012}%
  \BibitemOpen
  \bibfield  {author} {\bibinfo {author} {\bibfnamefont {G.}~\bibnamefont
  {Scalari}}, \bibinfo {author} {\bibfnamefont {C.}~\bibnamefont {Maissen}},
  \bibinfo {author} {\bibfnamefont {D.}~\bibnamefont {Tur{\v c}inkov{\'a}}},
  \bibinfo {author} {\bibfnamefont {D.}~\bibnamefont {Hagenm{\"u}ller}},
  \bibinfo {author} {\bibfnamefont {S.}~\bibnamefont {De~Liberato}}, \bibinfo
  {author} {\bibfnamefont {C.}~\bibnamefont {Ciuti}}, \bibinfo {author}
  {\bibfnamefont {C.}~\bibnamefont {Reichl}}, \bibinfo {author} {\bibfnamefont
  {D.}~\bibnamefont {Schuh}}, \bibinfo {author} {\bibfnamefont
  {W.}~\bibnamefont {Wegscheider}}, \bibinfo {author} {\bibfnamefont
  {M.}~\bibnamefont {Beck}}, \ and\ \bibinfo {author} {\bibfnamefont
  {J.}~\bibnamefont {Faist}},\ }\href {\doibase 10.1126/science.1216022}
  {\bibfield  {journal} {\bibinfo  {journal} {Science}\ }\textbf {\bibinfo
  {volume} {335}},\ \bibinfo {pages} {1323} (\bibinfo {year} {2012})},\ \Eprint
  {http://arxiv.org/abs/http://science.sciencemag.org/content/335/6074/1323.full.pdf}
  {http://science.sciencemag.org/content/335/6074/1323.full.pdf} \BibitemShut
  {NoStop}%
\bibitem [{\citenamefont {Keller}\ \emph {et~al.}(2018)\citenamefont {Keller},
  \citenamefont {Scalari}, \citenamefont {Appugliese}, \citenamefont {Maissen},
  \citenamefont {Haase}, \citenamefont {Failla}, \citenamefont {Myronov},
  \citenamefont {Leadley}, \citenamefont {Lloyd-Hughes}, \citenamefont
  {Nataf},\ and\ \citenamefont {Faist}}]{Keller2018}%
  \BibitemOpen
  \bibfield  {author} {\bibinfo {author} {\bibfnamefont {J.}~\bibnamefont
  {Keller}}, \bibinfo {author} {\bibfnamefont {G.}~\bibnamefont {Scalari}},
  \bibinfo {author} {\bibfnamefont {F.}~\bibnamefont {Appugliese}}, \bibinfo
  {author} {\bibfnamefont {C.}~\bibnamefont {Maissen}}, \bibinfo {author}
  {\bibfnamefont {J.}~\bibnamefont {Haase}}, \bibinfo {author} {\bibfnamefont
  {M.}~\bibnamefont {Failla}}, \bibinfo {author} {\bibfnamefont
  {M.}~\bibnamefont {Myronov}}, \bibinfo {author} {\bibfnamefont {D.~R.}\
  \bibnamefont {Leadley}}, \bibinfo {author} {\bibfnamefont {J.}~\bibnamefont
  {Lloyd-Hughes}}, \bibinfo {author} {\bibfnamefont {P.}~\bibnamefont {Nataf}},
  \ and\ \bibinfo {author} {\bibfnamefont {J.}~\bibnamefont {Faist}},\ }\href
  {http://arxiv.org/abs/1708.07773v1} {\bibfield  {journal} {\bibinfo
  {journal} {ArXiv e-prints}\ } (\bibinfo {year} {2018})},\ \Eprint
  {http://arxiv.org/abs/1708.07773} {arXiv:1708.07773} \BibitemShut {NoStop}%
\bibitem [{\citenamefont {Rza\ifmmode~\dot{z}\else \.{z}\fi{}ewski}\ \emph
  {et~al.}(1975)\citenamefont {Rza\ifmmode~\dot{z}\else \.{z}\fi{}ewski},
  \citenamefont {W\'odkiewicz},\ and\ \citenamefont {\ifmmode~\dot{Z}\else
  \.{Z}\fi{}akowicz}}]{polonais_1975}%
  \BibitemOpen
  \bibfield  {author} {\bibinfo {author} {\bibfnamefont {K.}~\bibnamefont
  {Rza\ifmmode~\dot{z}\else \.{z}\fi{}ewski}}, \bibinfo {author} {\bibfnamefont
  {K.}~\bibnamefont {W\'odkiewicz}}, \ and\ \bibinfo {author} {\bibfnamefont
  {W.}~\bibnamefont {\ifmmode~\dot{Z}\else \.{Z}\fi{}akowicz}},\ }\href
  {\doibase 10.1103/PhysRevLett.35.432} {\bibfield  {journal} {\bibinfo
  {journal} {Phys. Rev. Lett.}\ }\textbf {\bibinfo {volume} {35}},\ \bibinfo
  {pages} {432} (\bibinfo {year} {1975})}\BibitemShut {NoStop}%
\bibitem [{\citenamefont {Nataf}\ and\ \citenamefont
  {Ciuti}(2010{\natexlab{a}})}]{nataf_nogo_2010}%
  \BibitemOpen
  \bibfield  {author} {\bibinfo {author} {\bibfnamefont {P.}~\bibnamefont
  {Nataf}}\ and\ \bibinfo {author} {\bibfnamefont {C.}~\bibnamefont {Ciuti}},\
  }\href {\doibase 10.1038/ncomms1069} {\bibfield  {journal} {\bibinfo
  {journal} {Nature Communications}\ }\textbf {\bibinfo {volume} {1}},\
  \bibinfo {pages} {72} (\bibinfo {year} {2010}{\natexlab{a}})}\BibitemShut
  {NoStop}%
\bibitem [{\citenamefont {Todorov}\ and\ \citenamefont
  {Sirtori}(2012)}]{Todorov_2012}%
  \BibitemOpen
  \bibfield  {author} {\bibinfo {author} {\bibfnamefont {Y.}~\bibnamefont
  {Todorov}}\ and\ \bibinfo {author} {\bibfnamefont {C.}~\bibnamefont
  {Sirtori}},\ }\href {\doibase 10.1103/PhysRevB.85.045304} {\bibfield
  {journal} {\bibinfo  {journal} {Phys. Rev. B}\ }\textbf {\bibinfo {volume}
  {85}},\ \bibinfo {pages} {045304} (\bibinfo {year} {2012})}\BibitemShut
  {NoStop}%
\bibitem [{\citenamefont {Hayn}\ \emph {et~al.}(2012)\citenamefont {Hayn},
  \citenamefont {Emary},\ and\ \citenamefont {Brandes}}]{Hayn_2012}%
  \BibitemOpen
  \bibfield  {author} {\bibinfo {author} {\bibfnamefont {M.}~\bibnamefont
  {Hayn}}, \bibinfo {author} {\bibfnamefont {C.}~\bibnamefont {Emary}}, \ and\
  \bibinfo {author} {\bibfnamefont {T.}~\bibnamefont {Brandes}},\ }\href
  {\doibase 10.1103/PhysRevA.86.063822} {\bibfield  {journal} {\bibinfo
  {journal} {Phys. Rev. A}\ }\textbf {\bibinfo {volume} {86}},\ \bibinfo
  {pages} {063822} (\bibinfo {year} {2012})}\BibitemShut {NoStop}%
\bibitem [{\citenamefont {Chirolli}\ \emph {et~al.}(2012)\citenamefont
  {Chirolli}, \citenamefont {Polini}, \citenamefont {Giovannetti},\ and\
  \citenamefont {MacDonald}}]{polini_2012}%
  \BibitemOpen
  \bibfield  {author} {\bibinfo {author} {\bibfnamefont {L.}~\bibnamefont
  {Chirolli}}, \bibinfo {author} {\bibfnamefont {M.}~\bibnamefont {Polini}},
  \bibinfo {author} {\bibfnamefont {V.}~\bibnamefont {Giovannetti}}, \ and\
  \bibinfo {author} {\bibfnamefont {A.~H.}\ \bibnamefont {MacDonald}},\ }\href
  {\doibase 10.1103/PhysRevLett.109.267404} {\bibfield  {journal} {\bibinfo
  {journal} {Phys. Rev. Lett.}\ }\textbf {\bibinfo {volume} {109}},\ \bibinfo
  {pages} {267404} (\bibinfo {year} {2012})}\BibitemShut {NoStop}%
\bibitem [{\citenamefont {Bamba}\ and\ \citenamefont
  {Ogawa}(2014)}]{Bamba_2014}%
  \BibitemOpen
  \bibfield  {author} {\bibinfo {author} {\bibfnamefont {M.}~\bibnamefont
  {Bamba}}\ and\ \bibinfo {author} {\bibfnamefont {T.}~\bibnamefont {Ogawa}},\
  }\href {\doibase 10.1103/PhysRevA.90.063825} {\bibfield  {journal} {\bibinfo
  {journal} {Phys. Rev. A}\ }\textbf {\bibinfo {volume} {90}},\ \bibinfo
  {pages} {063825} (\bibinfo {year} {2014})}\BibitemShut {NoStop}%
\bibitem [{\citenamefont {Rousseau}\ and\ \citenamefont
  {Felbacq}(2017)}]{rousseau_2017}%
  \BibitemOpen
  \bibfield  {author} {\bibinfo {author} {\bibfnamefont {E.}~\bibnamefont
  {Rousseau}}\ and\ \bibinfo {author} {\bibfnamefont {D.}~\bibnamefont
  {Felbacq}},\ }\href {\doibase 110.1038/s41598-017-11076-5} {\bibfield
  {journal} {\bibinfo  {journal} {Scientific Reports}\ }\textbf {\bibinfo
  {volume} {7}},\ \bibinfo {pages} {11115} (\bibinfo {year}
  {2017})}\BibitemShut {NoStop}%
\bibitem [{\citenamefont {{Andolina}}\ \emph {et~al.}(2019)\citenamefont
  {{Andolina}}, \citenamefont {{Pellegrino}}, \citenamefont {{Giovannetti}},
  \citenamefont {{MacDonald}},\ and\ \citenamefont {{Polini}}}]{andolina_nogo}%
  \BibitemOpen
  \bibfield  {author} {\bibinfo {author} {\bibfnamefont {G.~M.}\ \bibnamefont
  {{Andolina}}}, \bibinfo {author} {\bibfnamefont {F.~M.~D.}\ \bibnamefont
  {{Pellegrino}}}, \bibinfo {author} {\bibfnamefont {V.}~\bibnamefont
  {{Giovannetti}}}, \bibinfo {author} {\bibfnamefont {A.~H.}\ \bibnamefont
  {{MacDonald}}}, \ and\ \bibinfo {author} {\bibfnamefont {M.}~\bibnamefont
  {{Polini}}},\ }\href@noop {} {\bibfield  {journal} {\bibinfo  {journal}
  {arXiv e-prints}\ ,\ \bibinfo {eid} {arXiv:1905.11227}} (\bibinfo {year}
  {2019})},\ \Eprint {http://arxiv.org/abs/1905.11227} {arXiv:1905.11227
  [cond-mat.mes-hall]} \BibitemShut {NoStop}%
\bibitem [{\citenamefont {Dimer}\ \emph {et~al.}(2007)\citenamefont {Dimer},
  \citenamefont {Estienne}, \citenamefont {Parkins},\ and\ \citenamefont
  {Carmichael}}]{Carmichael_PRA_2007}%
  \BibitemOpen
  \bibfield  {author} {\bibinfo {author} {\bibfnamefont {F.}~\bibnamefont
  {Dimer}}, \bibinfo {author} {\bibfnamefont {B.}~\bibnamefont {Estienne}},
  \bibinfo {author} {\bibfnamefont {A.~S.}\ \bibnamefont {Parkins}}, \ and\
  \bibinfo {author} {\bibfnamefont {H.~J.}\ \bibnamefont {Carmichael}},\ }\href
  {\doibase 10.1103/PhysRevA.75.013804} {\bibfield  {journal} {\bibinfo
  {journal} {Phys. Rev. A}\ }\textbf {\bibinfo {volume} {75}},\ \bibinfo
  {pages} {013804} (\bibinfo {year} {2007})}\BibitemShut {NoStop}%
\bibitem [{\citenamefont {Damanet}\ \emph {et~al.}(2019)\citenamefont
  {Damanet}, \citenamefont {Daley},\ and\ \citenamefont
  {Keeling}}]{Keeling_PRA_2019}%
  \BibitemOpen
  \bibfield  {author} {\bibinfo {author} {\bibfnamefont {F.~m.~c.}\
  \bibnamefont {Damanet}}, \bibinfo {author} {\bibfnamefont {A.~J.}\
  \bibnamefont {Daley}}, \ and\ \bibinfo {author} {\bibfnamefont
  {J.}~\bibnamefont {Keeling}},\ }\href {\doibase 10.1103/PhysRevA.99.033845}
  {\bibfield  {journal} {\bibinfo  {journal} {Phys. Rev. A}\ }\textbf {\bibinfo
  {volume} {99}},\ \bibinfo {pages} {033845} (\bibinfo {year}
  {2019})}\BibitemShut {NoStop}%
\bibitem [{\citenamefont {Baumann}\ \emph {et~al.}(2010)\citenamefont
  {Baumann}, \citenamefont {Guerlin}, \citenamefont {Brennecke},\ and\
  \citenamefont {Esslinger}}]{esslinger_2010}%
  \BibitemOpen
  \bibfield  {author} {\bibinfo {author} {\bibfnamefont {K.}~\bibnamefont
  {Baumann}}, \bibinfo {author} {\bibfnamefont {C.}~\bibnamefont {Guerlin}},
  \bibinfo {author} {\bibfnamefont {F.}~\bibnamefont {Brennecke}}, \ and\
  \bibinfo {author} {\bibfnamefont {T.}~\bibnamefont {Esslinger}},\ }\href
  {\doibase 10.1038/nature09009} {\bibfield  {journal} {\bibinfo  {journal}
  {Nature}\ }\textbf {\bibinfo {volume} {464}},\ \bibinfo {pages} {1301}
  (\bibinfo {year} {2010})}\BibitemShut {NoStop}%
\bibitem [{\citenamefont {Knight}\ \emph {et~al.}(1978)\citenamefont {Knight},
  \citenamefont {Aharonov},\ and\ \citenamefont {Hsieh}}]{Knight_PRA}%
  \BibitemOpen
  \bibfield  {author} {\bibinfo {author} {\bibfnamefont {J.~M.}\ \bibnamefont
  {Knight}}, \bibinfo {author} {\bibfnamefont {Y.}~\bibnamefont {Aharonov}}, \
  and\ \bibinfo {author} {\bibfnamefont {G.~T.~C.}\ \bibnamefont {Hsieh}},\
  }\href {\doibase 10.1103/PhysRevA.17.1454} {\bibfield  {journal} {\bibinfo
  {journal} {Phys. Rev. A}\ }\textbf {\bibinfo {volume} {17}},\ \bibinfo
  {pages} {1454} (\bibinfo {year} {1978})}\BibitemShut {NoStop}%
\bibitem [{\citenamefont {Nataf}\ and\ \citenamefont
  {Ciuti}(2010{\natexlab{b}})}]{nataf_2010}%
  \BibitemOpen
  \bibfield  {author} {\bibinfo {author} {\bibfnamefont {P.}~\bibnamefont
  {Nataf}}\ and\ \bibinfo {author} {\bibfnamefont {C.}~\bibnamefont {Ciuti}},\
  }\href {\doibase 10.1103/PhysRevLett.104.023601} {\bibfield  {journal}
  {\bibinfo  {journal} {Phys. Rev. Lett.}\ }\textbf {\bibinfo {volume} {104}},\
  \bibinfo {pages} {023601} (\bibinfo {year} {2010}{\natexlab{b}})}\BibitemShut
  {NoStop}%
\bibitem [{\citenamefont {Nataf}\ and\ \citenamefont
  {Ciuti}(2011)}]{nataf_2011}%
  \BibitemOpen
  \bibfield  {author} {\bibinfo {author} {\bibfnamefont {P.}~\bibnamefont
  {Nataf}}\ and\ \bibinfo {author} {\bibfnamefont {C.}~\bibnamefont {Ciuti}},\
  }\href {\doibase 10.1103/PhysRevLett.107.190402} {\bibfield  {journal}
  {\bibinfo  {journal} {Phys. Rev. Lett.}\ }\textbf {\bibinfo {volume} {107}},\
  \bibinfo {pages} {190402} (\bibinfo {year} {2011})}\BibitemShut {NoStop}%
\bibitem [{\citenamefont {Devoret}\ \emph {et~al.}(2007)\citenamefont
  {Devoret}, \citenamefont {Girvin},\ and\ \citenamefont
  {Schoelkopf}}]{devoret_2007}%
  \BibitemOpen
  \bibfield  {author} {\bibinfo {author} {\bibfnamefont {M.}~\bibnamefont
  {Devoret}}, \bibinfo {author} {\bibfnamefont {S.}~\bibnamefont {Girvin}}, \
  and\ \bibinfo {author} {\bibfnamefont {R.}~\bibnamefont {Schoelkopf}},\
  }\href {\doibase 10.1002/andp.200710261} {\bibfield  {journal} {\bibinfo
  {journal} {Annalen der Physik}\ }\textbf {\bibinfo {volume} {16}},\ \bibinfo
  {pages} {767} (\bibinfo {year} {2007})}\BibitemShut {NoStop}%
\bibitem [{\citenamefont {Keeling}(2007)}]{Keeling_2007}%
  \BibitemOpen
  \bibfield  {author} {\bibinfo {author} {\bibfnamefont {J.}~\bibnamefont
  {Keeling}},\ }\href {http://stacks.iop.org/0953-8984/19/i=29/a=295213}
  {\bibfield  {journal} {\bibinfo  {journal} {Journal of Physics: Condensed
  Matter}\ }\textbf {\bibinfo {volume} {19}},\ \bibinfo {pages} {295213}
  (\bibinfo {year} {2007})}\BibitemShut {NoStop}%
\bibitem [{\citenamefont {Vukics}\ \emph {et~al.}(2015)\citenamefont {Vukics},
  \citenamefont {Grie\ss{}er},\ and\ \citenamefont {Domokos}}]{Vukics_2015}%
  \BibitemOpen
  \bibfield  {author} {\bibinfo {author} {\bibfnamefont {A.}~\bibnamefont
  {Vukics}}, \bibinfo {author} {\bibfnamefont {T.}~\bibnamefont {Grie\ss{}er}},
  \ and\ \bibinfo {author} {\bibfnamefont {P.}~\bibnamefont {Domokos}},\ }\href
  {\doibase 10.1103/PhysRevA.92.043835} {\bibfield  {journal} {\bibinfo
  {journal} {Phys. Rev. A}\ }\textbf {\bibinfo {volume} {92}},\ \bibinfo
  {pages} {043835} (\bibinfo {year} {2015})}\BibitemShut {NoStop}%
\bibitem [{\citenamefont {Pellegrino}\ \emph {et~al.}(2016)\citenamefont
  {Pellegrino}, \citenamefont {Giovannetti}, \citenamefont {MacDonald},\ and\
  \citenamefont {Polini}}]{mac_donald_ncomm}%
  \BibitemOpen
  \bibfield  {author} {\bibinfo {author} {\bibfnamefont {F.~M.~D.}\
  \bibnamefont {Pellegrino}}, \bibinfo {author} {\bibfnamefont
  {V.}~\bibnamefont {Giovannetti}}, \bibinfo {author} {\bibfnamefont {A.~H.}\
  \bibnamefont {MacDonald}}, \ and\ \bibinfo {author} {\bibfnamefont
  {M.}~\bibnamefont {Polini}},\ }\href {\doibase 10.1038/ncomms13355}
  {\bibfield  {journal} {\bibinfo  {journal} {Nature Communications}\ }\textbf
  {\bibinfo {volume} {7}},\ \bibinfo {pages} {13355} (\bibinfo {year}
  {2016})}\BibitemShut {NoStop}%
\bibitem [{\citenamefont {Mazza}\ and\ \citenamefont
  {Georges}(2019)}]{Georges_PRL_2019}%
  \BibitemOpen
  \bibfield  {author} {\bibinfo {author} {\bibfnamefont {G.}~\bibnamefont
  {Mazza}}\ and\ \bibinfo {author} {\bibfnamefont {A.}~\bibnamefont
  {Georges}},\ }\href {\doibase 10.1103/PhysRevLett.122.017401} {\bibfield
  {journal} {\bibinfo  {journal} {Phys. Rev. Lett.}\ }\textbf {\bibinfo
  {volume} {122}},\ \bibinfo {pages} {017401} (\bibinfo {year} {2019})},\
  \bibinfo {note} {{the mistake of this paper is to conclude that the excitonic
  instability is accompanied by spontaneous generation of a static and
  spatially uniform average value of the vector potential}}\BibitemShut
  {NoStop}%
\bibitem [{\citenamefont {Shoenberg}(1984)}]{ShoenbergBook}%
  \BibitemOpen
  \bibfield  {author} {\bibinfo {author} {\bibfnamefont {D.}~\bibnamefont
  {Shoenberg}},\ }\href@noop {} {\emph {\bibinfo {title} {{Magnetic
  Oscillations in Metals}}}}\ (\bibinfo  {publisher} {{Cambridge University
  Press}},\ \bibinfo {year} {1984})\BibitemShut {NoStop}%
\bibitem [{\citenamefont {Weisbuch}\ \emph {et~al.}(1992)\citenamefont
  {Weisbuch}, \citenamefont {Nishioka}, \citenamefont {Ishikawa},\ and\
  \citenamefont {Arakawa}}]{Weisbuch_1992}%
  \BibitemOpen
  \bibfield  {author} {\bibinfo {author} {\bibfnamefont {C.}~\bibnamefont
  {Weisbuch}}, \bibinfo {author} {\bibfnamefont {M.}~\bibnamefont {Nishioka}},
  \bibinfo {author} {\bibfnamefont {A.}~\bibnamefont {Ishikawa}}, \ and\
  \bibinfo {author} {\bibfnamefont {Y.}~\bibnamefont {Arakawa}},\ }\href
  {\doibase 10.1103/PhysRevLett.69.3314} {\bibfield  {journal} {\bibinfo
  {journal} {Phys. Rev. Lett.}\ }\textbf {\bibinfo {volume} {69}},\ \bibinfo
  {pages} {3314} (\bibinfo {year} {1992})}\BibitemShut {NoStop}%
\bibitem [{\citenamefont {{Bychkov}}\ and\ \citenamefont
  {{Rashba}}(1984)}]{rashba_1984}%
  \BibitemOpen
  \bibfield  {author} {\bibinfo {author} {\bibfnamefont {Y.~A.}\ \bibnamefont
  {{Bychkov}}}\ and\ \bibinfo {author} {\bibfnamefont {{\'E}.~I.}\ \bibnamefont
  {{Rashba}}},\ }\href@noop {} {\bibfield  {journal} {\bibinfo  {journal} {JETP
  Letters}\ }\textbf {\bibinfo {volume} {39}},\ \bibinfo {pages} {78} (\bibinfo
  {year} {1984})}\BibitemShut {NoStop}%
\bibitem [{\citenamefont {Becker}\ \emph {et~al.}(2010)\citenamefont {Becker},
  \citenamefont {Liebmann}, \citenamefont {Mashoff}, \citenamefont {Pratzer},\
  and\ \citenamefont {Morgenstern}}]{Morgenstern_2010}%
  \BibitemOpen
  \bibfield  {author} {\bibinfo {author} {\bibfnamefont {S.}~\bibnamefont
  {Becker}}, \bibinfo {author} {\bibfnamefont {M.}~\bibnamefont {Liebmann}},
  \bibinfo {author} {\bibfnamefont {T.}~\bibnamefont {Mashoff}}, \bibinfo
  {author} {\bibfnamefont {M.}~\bibnamefont {Pratzer}}, \ and\ \bibinfo
  {author} {\bibfnamefont {M.}~\bibnamefont {Morgenstern}},\ }\href {\doibase
  10.1103/PhysRevB.81.155308} {\bibfield  {journal} {\bibinfo  {journal} {Phys.
  Rev. B}\ }\textbf {\bibinfo {volume} {81}},\ \bibinfo {pages} {155308}
  (\bibinfo {year} {2010})}\BibitemShut {NoStop}%
\bibitem [{\citenamefont {Hernang\'omez-P\'erez}\ \emph
  {et~al.}(2013)\citenamefont {Hernang\'omez-P\'erez}, \citenamefont {Ulrich},
  \citenamefont {Florens},\ and\ \citenamefont {Champel}}]{Champel2013}%
  \BibitemOpen
  \bibfield  {author} {\bibinfo {author} {\bibfnamefont {D.}~\bibnamefont
  {Hernang\'omez-P\'erez}}, \bibinfo {author} {\bibfnamefont {J.}~\bibnamefont
  {Ulrich}}, \bibinfo {author} {\bibfnamefont {S.}~\bibnamefont {Florens}}, \
  and\ \bibinfo {author} {\bibfnamefont {T.}~\bibnamefont {Champel}},\ }\href
  {\doibase 10.1103/PhysRevB.88.245433} {\bibfield  {journal} {\bibinfo
  {journal} {Phys. Rev. B}\ }\textbf {\bibinfo {volume} {88}},\ \bibinfo
  {pages} {245433} (\bibinfo {year} {2013})}\BibitemShut {NoStop}%
\bibitem [{\citenamefont {Kakazu}\ and\ \citenamefont
  {Kim}(1994)}]{kakazu_kim}%
  \BibitemOpen
  \bibfield  {author} {\bibinfo {author} {\bibfnamefont {K.}~\bibnamefont
  {Kakazu}}\ and\ \bibinfo {author} {\bibfnamefont {Y.~S.}\ \bibnamefont
  {Kim}},\ }\href {\doibase 10.1103/PhysRevA.50.1830} {\bibfield  {journal}
  {\bibinfo  {journal} {Phys. Rev. A}\ }\textbf {\bibinfo {volume} {50}},\
  \bibinfo {pages} {1830} (\bibinfo {year} {1994})}\BibitemShut {NoStop}%
\bibitem [{\citenamefont {Hagenm\"uller}\ \emph {et~al.}(2010)\citenamefont
  {Hagenm\"uller}, \citenamefont {De~Liberato},\ and\ \citenamefont
  {Ciuti}}]{Hagenmuller_2010}%
  \BibitemOpen
  \bibfield  {author} {\bibinfo {author} {\bibfnamefont {D.}~\bibnamefont
  {Hagenm\"uller}}, \bibinfo {author} {\bibfnamefont {S.}~\bibnamefont
  {De~Liberato}}, \ and\ \bibinfo {author} {\bibfnamefont {C.}~\bibnamefont
  {Ciuti}},\ }\href {\doibase 10.1103/PhysRevB.81.235303} {\bibfield  {journal}
  {\bibinfo  {journal} {Phys. Rev. B}\ }\textbf {\bibinfo {volume} {81}},\
  \bibinfo {pages} {235303} (\bibinfo {year} {2010})}\BibitemShut {NoStop}%
\bibitem [{\citenamefont {Nataf}\ \emph {et~al.}(2012)\citenamefont {Nataf},
  \citenamefont {Baksic},\ and\ \citenamefont {Ciuti}}]{Nataf_2012}%
  \BibitemOpen
  \bibfield  {author} {\bibinfo {author} {\bibfnamefont {P.}~\bibnamefont
  {Nataf}}, \bibinfo {author} {\bibfnamefont {A.}~\bibnamefont {Baksic}}, \
  and\ \bibinfo {author} {\bibfnamefont {C.}~\bibnamefont {Ciuti}},\ }\href
  {\doibase 10.1103/PhysRevA.86.013832} {\bibfield  {journal} {\bibinfo
  {journal} {Phys. Rev. A}\ }\textbf {\bibinfo {volume} {86}},\ \bibinfo
  {pages} {013832} (\bibinfo {year} {2012})}\BibitemShut {NoStop}%
\bibitem [{\citenamefont {Baksic}\ and\ \citenamefont
  {Ciuti}(2014)}]{Baksic_2014}%
  \BibitemOpen
  \bibfield  {author} {\bibinfo {author} {\bibfnamefont {A.}~\bibnamefont
  {Baksic}}\ and\ \bibinfo {author} {\bibfnamefont {C.}~\bibnamefont {Ciuti}},\
  }\href {\doibase 10.1103/PhysRevLett.112.173601} {\bibfield  {journal}
  {\bibinfo  {journal} {Phys. Rev. Lett.}\ }\textbf {\bibinfo {volume} {112}},\
  \bibinfo {pages} {173601} (\bibinfo {year} {2014})}\BibitemShut {NoStop}%
\bibitem [{\citenamefont {Champel}\ and\ \citenamefont
  {Florens}(2009)}]{Champel2009}%
  \BibitemOpen
  \bibfield  {author} {\bibinfo {author} {\bibfnamefont {T.}~\bibnamefont
  {Champel}}\ and\ \bibinfo {author} {\bibfnamefont {S.}~\bibnamefont
  {Florens}},\ }\href {\doibase 10.1103/PhysRevB.80.161311} {\bibfield
  {journal} {\bibinfo  {journal} {Phys. Rev. B}\ }\textbf {\bibinfo {volume}
  {80}},\ \bibinfo {pages} {161311} (\bibinfo {year} {2009})}\BibitemShut
  {NoStop}%
\bibitem [{\citenamefont {Luttinger}(1956)}]{luttinger}%
  \BibitemOpen
  \bibfield  {author} {\bibinfo {author} {\bibfnamefont {J.~M.}\ \bibnamefont
  {Luttinger}},\ }\href {\doibase 10.1103/PhysRev.102.1030} {\bibfield
  {journal} {\bibinfo  {journal} {Phys. Rev.}\ }\textbf {\bibinfo {volume}
  {102}},\ \bibinfo {pages} {1030} (\bibinfo {year} {1956})}\BibitemShut
  {NoStop}%
\bibitem [{\citenamefont {Winkler}\ \emph {et~al.}(1996)\citenamefont
  {Winkler}, \citenamefont {Merkler}, \citenamefont {Darnhofer},\ and\
  \citenamefont {R\"ossler}}]{Winkler_1995}%
  \BibitemOpen
  \bibfield  {author} {\bibinfo {author} {\bibfnamefont {R.}~\bibnamefont
  {Winkler}}, \bibinfo {author} {\bibfnamefont {M.}~\bibnamefont {Merkler}},
  \bibinfo {author} {\bibfnamefont {T.}~\bibnamefont {Darnhofer}}, \ and\
  \bibinfo {author} {\bibfnamefont {U.}~\bibnamefont {R\"ossler}},\ }\href
  {\doibase 10.1103/PhysRevB.53.10858} {\bibfield  {journal} {\bibinfo
  {journal} {Phys. Rev. B}\ }\textbf {\bibinfo {volume} {53}},\ \bibinfo
  {pages} {10858} (\bibinfo {year} {1996})}\BibitemShut {NoStop}%
\bibitem [{\citenamefont {Winkler}(2000)}]{Winkler_2000}%
  \BibitemOpen
  \bibfield  {author} {\bibinfo {author} {\bibfnamefont {R.}~\bibnamefont
  {Winkler}},\ }\href {\doibase 10.1103/PhysRevB.62.4245} {\bibfield  {journal}
  {\bibinfo  {journal} {Phys. Rev. B}\ }\textbf {\bibinfo {volume} {62}},\
  \bibinfo {pages} {4245} (\bibinfo {year} {2000})}\BibitemShut {NoStop}%
\bibitem [{\citenamefont {Moriya}\ \emph {et~al.}(2014)\citenamefont {Moriya},
  \citenamefont {Sawano}, \citenamefont {Hoshi}, \citenamefont {Masubuchi},
  \citenamefont {Shiraki}, \citenamefont {Wild}, \citenamefont {Neumann},
  \citenamefont {Abstreiter}, \citenamefont {Bougeard}, \citenamefont {Koga},\
  and\ \citenamefont {Machida}}]{Moriya_PRL}%
  \BibitemOpen
  \bibfield  {author} {\bibinfo {author} {\bibfnamefont {R.}~\bibnamefont
  {Moriya}}, \bibinfo {author} {\bibfnamefont {K.}~\bibnamefont {Sawano}},
  \bibinfo {author} {\bibfnamefont {Y.}~\bibnamefont {Hoshi}}, \bibinfo
  {author} {\bibfnamefont {S.}~\bibnamefont {Masubuchi}}, \bibinfo {author}
  {\bibfnamefont {Y.}~\bibnamefont {Shiraki}}, \bibinfo {author} {\bibfnamefont
  {A.}~\bibnamefont {Wild}}, \bibinfo {author} {\bibfnamefont {C.}~\bibnamefont
  {Neumann}}, \bibinfo {author} {\bibfnamefont {G.}~\bibnamefont {Abstreiter}},
  \bibinfo {author} {\bibfnamefont {D.}~\bibnamefont {Bougeard}}, \bibinfo
  {author} {\bibfnamefont {T.}~\bibnamefont {Koga}}, \ and\ \bibinfo {author}
  {\bibfnamefont {T.}~\bibnamefont {Machida}},\ }\href {\doibase
  10.1103/PhysRevLett.113.086601} {\bibfield  {journal} {\bibinfo  {journal}
  {Phys. Rev. Lett.}\ }\textbf {\bibinfo {volume} {113}},\ \bibinfo {pages}
  {086601} (\bibinfo {year} {2014})}\BibitemShut {NoStop}%
\bibitem [{\citenamefont {Maissen}\ \emph {et~al.}(2014)\citenamefont
  {Maissen}, \citenamefont {Scalari}, \citenamefont {Valmorra}, \citenamefont
  {Beck}, \citenamefont {Faist}, \citenamefont {Cibella}, \citenamefont
  {Leoni}, \citenamefont {Reichl}, \citenamefont {Charpentier},\ and\
  \citenamefont {Wegscheider}}]{Maissen_2014}%
  \BibitemOpen
  \bibfield  {author} {\bibinfo {author} {\bibfnamefont {C.}~\bibnamefont
  {Maissen}}, \bibinfo {author} {\bibfnamefont {G.}~\bibnamefont {Scalari}},
  \bibinfo {author} {\bibfnamefont {F.}~\bibnamefont {Valmorra}}, \bibinfo
  {author} {\bibfnamefont {M.}~\bibnamefont {Beck}}, \bibinfo {author}
  {\bibfnamefont {J.}~\bibnamefont {Faist}}, \bibinfo {author} {\bibfnamefont
  {S.}~\bibnamefont {Cibella}}, \bibinfo {author} {\bibfnamefont
  {R.}~\bibnamefont {Leoni}}, \bibinfo {author} {\bibfnamefont
  {C.}~\bibnamefont {Reichl}}, \bibinfo {author} {\bibfnamefont
  {C.}~\bibnamefont {Charpentier}}, \ and\ \bibinfo {author} {\bibfnamefont
  {W.}~\bibnamefont {Wegscheider}},\ }\href {\doibase
  10.1103/PhysRevB.90.205309} {\bibfield  {journal} {\bibinfo  {journal} {Phys.
  Rev. B}\ }\textbf {\bibinfo {volume} {90}},\ \bibinfo {pages} {205309}
  (\bibinfo {year} {2014})}\BibitemShut {NoStop}%
\end{thebibliography}%


\end{document}